\documentclass[journal=ancac3, manuscript=article]{achemso}
\setkeys{acs}{maxauthors=20, etalmode=truncate}
\usepackage[version=3]{mhchem} 
\usepackage[english]{babel}
\usepackage[utf8]{inputenc} % proper recognition of "ü", "ö" and 
\usepackage{siunitx} % proper unit formatting
\usepackage{hyperref} % needed for \sfigref
\usepackage{xcolor} % needed for highlighting changes
\usepackage{soul} % needed for highlighting changes - strikethrough
\usepackage{changepage}
\usepackage{gensymb}
\usepackage{amsmath}
\usepackage{graphicx}
\usepackage{ulem}

\definecolor{delftdark}{rgb}{0, .5, .7}

\title{\Large Ultimately-scaled electrodes for contacting individual atomically-precise graphene nanoribbons}

\author{Jian Zhang}
\affiliation{\small Transport at Nanoscale Interfaces Laboratory, Empa, Swiss Federal Laboratories for Materials Science and Technology, 8600 Dübendorf, Switzerland}
\email{jian.zhang@empa.ch}

\author{Liu Qian}
\affiliation{\small College of Chemistry and Molecular Engineering, Peking University, 100871 Beijing, China}

\author{Gabriela Borin Barin}
\affiliation{\small nanotech@surfaces Laboratory, Empa, Swiss Federal Laboratories for Materials Science and Technology, 8600 Dübendorf, Switzerland}

\author{Abdalghani H.S. Daaoub}
\affiliation{\small School of Engineering, University of Warwick, Coventry CV4 7AL, United Kingdom}

\author{Peipei Chen}
\affiliation{\small Nanofabrication Laboratory, National Center for Nanoscience and Technology, 100190 Beijing, China}

\author{Klaus M\"ullen}
\affiliation{\small Max Planck Institute for Polymer Research, 55128 Mainz, Germany}

\author{Sara Sangtarash}
\affiliation{\small School of Engineering, University of Warwick, Coventry CV4 7AL, United Kingdom}

\author{Pascal Ruffieux}
\affiliation{\small nanotech@surfaces Laboratory, Empa, Swiss Federal Laboratories for Materials Science and Technology, 8600 Dübendorf, Switzerland}

\author{Roman Fasel}
\affiliation{\small nanotech@surfaces Laboratory, Empa, Swiss Federal Laboratories for Materials Science and Technology, 8600 Dübendorf, Switzerland}
\alsoaffiliation{\small Department of Chemistry, Biochemistry and Pharmaceutical Sciences, University of Bern, 3012 Bern, Switzerland}

\author{Hatef Sadeghi}
\affiliation{\small School of Engineering, University of Warwick, Coventry CV4 7AL, United Kingdom}
\email{hatef.sadeghi@warwick.ac.uk}

\author{Jin Zhang}
\affiliation{\small College of Chemistry and Molecular Engineering, Peking University, 100871 Beijing, China}

\author{Michel Calame}
\affiliation{\small Transport at Nanoscale Interfaces Laboratory, Empa, Swiss Federal Laboratories for Materials Science and Technology, 8600 Dübendorf, Switzerland}
\alsoaffiliation{\small Department of Physics, University of Basel, 4056 Basel, Switzerland}
\alsoaffiliation{\small Swiss Nanoscience Institute, University of Basel, 4056 Basel, Switzerland}
\email{michel.calame@empa.ch}

\author{Mickael L. Perrin}
\affiliation{\small Transport at Nanoscale Interfaces Laboratory, Empa, Swiss Federal Laboratories for Materials Science and Technology, 8600 Dübendorf, Switzerland}
\alsoaffiliation{Department of Information Technology and Electrical Engineering, ETH Zurich, 8092 Zurich, Switzerland}
\email{mickael.perrin@empa.ch}

\begin{document}

\newpage

\textbf{Bottom-up synthesized graphene nanoribbons (GNRs) are quantum materials that can be structured with atomic precision, providing unprecedented control over their physical properties. Accessing the intrinsic functionality of GNRs for quantum technology applications requires the manipulation of single charges, spins, or photons at the level of an individual GNR. However, experimentally, contacting individual GNRs remains challenging due to their nanometer-sized width and length as well as their high density on the metallic growth substrate. Here, we demonstrate the contacting and electrical characterization of individual GNRs in a multi-gate device architecture using single-walled carbon nanotubes (SWNTs) as ultimately-scaled electrodes. The GNR-SWNT devices exhibit well-defined quantum transport phenomena, including Coulomb blockade, excited states, and Franck-Condon blockade, all characteristics pointing towards the contacting of an individual GNR. Combined with the multi-gate architecture, this contacting method opens a road for the integration of GNRs in quantum devices to exploit their topologically trivial and non-trivial nature.} \\

\begin{figure}
\begin{center}
	  \includegraphics[width=0.99\textwidth]{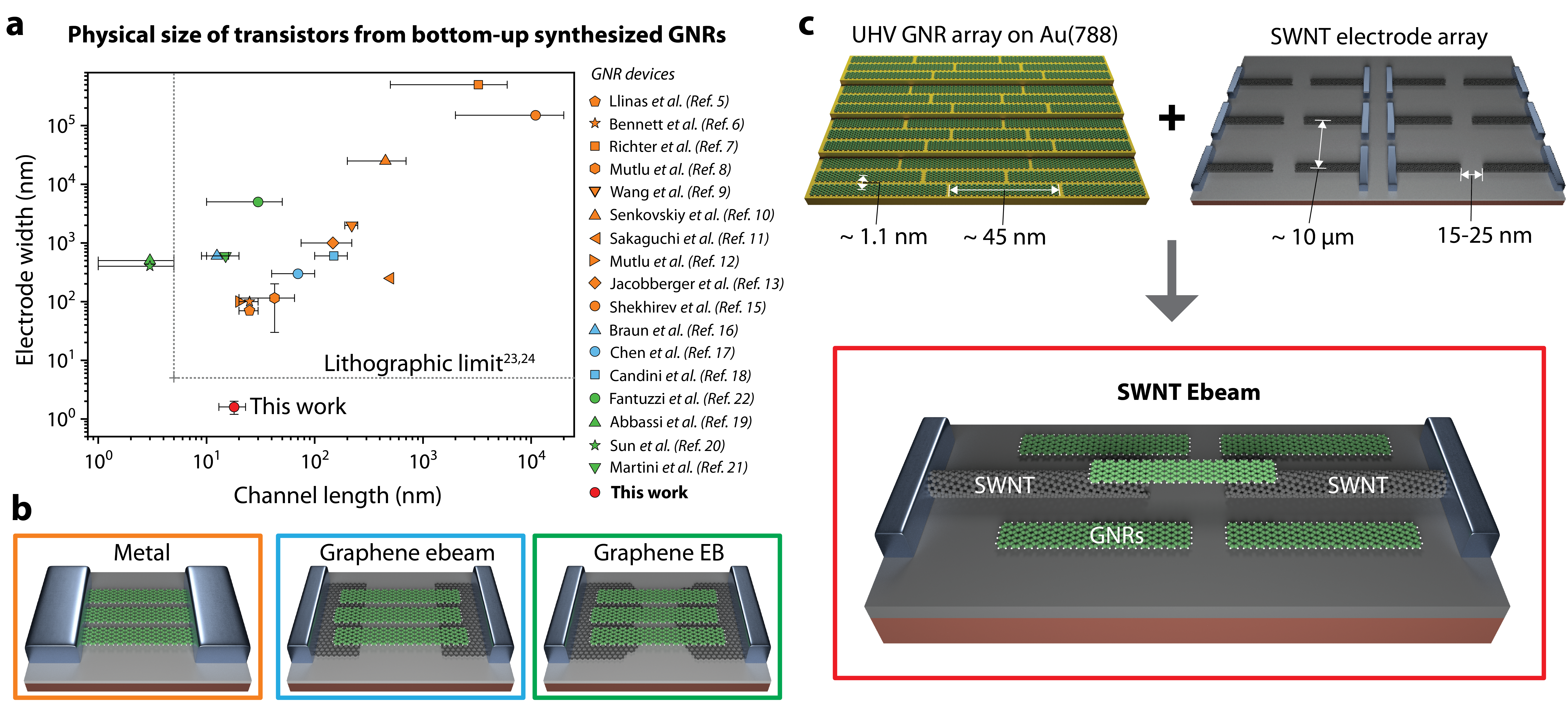} 
\end{center}
\caption{\textbf{Size scaling in bottom-up GNR based transistors with various geometries.} 
(\textbf{a}) Comparison of the physical size of transistors from GNRs with different contact strategies: Metal electrodes\citep{llinas2017short,bennett2013bottom,richter2020charge,mutlu2021short,wang2021towards,senkovskiy2021tunneling,sakaguchi2017homochiral,mutlu2021transfer,jacobberger2015direct,way2018seed,shekhirev2017interfacial} (Brown), e-beam lithographic graphene electrodes\citep{braun2021optimized,chen2016synthesis,candini2017high} (blue), electrical breakdown (EB) formed graphene electrodes\citep{el2020controlled,sun2020massive,martini2019structure,fantuzzi2016fabrication} (Green), and e-beam lithographic SWNT electrode (Red, this work). Square symbols represent surface-polymerized GNRs in UHV; Triangle symbols represent solution-polymerized GNRs; Circle symbols represent CVD synthesized GNRs.
    (\textbf{b}) Transistor schematics of typical bottom-up GNR transistors with metal electrodes (left), e-beam graphene electrodes (middle) and EB graphene electrodes (right).
    (\textbf{c}) Ultimately-scaled SWNT electrodes for contacting bottom-up graphene nanoribbons. Top-left: Schematic of the UHV synthesized GNR array parallel to the Au(788) terraces. Top-right: Schematic of the parallel SWNT electrode array on SiO2 substrate. Bottom: Schematic of a single-GNR based transistor with SWNTs as ultimately-scaled electrodes.
    }
    \label{fig:intro}
\end{figure}

Graphene nanoribbons (GNR) are an ultra-tunable class of designer quantum material with a wide range of electronic, magnetic and optical properties, ranging from ultra-low to large bandgap semiconductors and single-photon emitters, to the exhibition of spin-polarized and topologically protected states.\citep{Ruffieux2016surfaceSynthesisGraphene,cai2010atomically,Groning2018EngineeringRobustTopological,Rizzo2018TopologicalBandEngineering} Exploiting these appealing properties for quantum technology purposes requires both control over their chemical structure, as well as their integration into device architectures. The ability to integrate and contact an individual GNR opens up potential applications for their use as, for example, semiconducting quantum dots for trapping individual charges and the associated spins for the realization of charge or spin qubits, as well as single-photon emitters. However, contacting an individual GNR is a very challenging task. To date, bottom-up synthesized GNRs have been contacted using different approaches (see overview in Fig.~\ref{fig:intro}a-b), with the electrode material either a noble metal (gold, platinum, or palladium)\citep{llinas2017short,bennett2013bottom,richter2020charge,mutlu2021short,wang2021towards,senkovskiy2021tunneling,sakaguchi2017homochiral,mutlu2021transfer,jacobberger2015direct,way2018seed,shekhirev2017interfacial} or graphene\citep{braun2021optimized,chen2016synthesis,candini2017high,el2020controlled,sun2020massive,martini2019structure,fantuzzi2016fabrication}. Metallic electrodes are widely used, with the metal deposited either before or after the GNR transfer, referred to as ``GNR-last'' and ``GNR-first'', respectively, with ``GNR-first'' being the method of choice for ultra-short channel lenghts.\citep{llinas2017short,bennett2013bottom,richter2020charge,mutlu2021short,wang2021towards,senkovskiy2021tunneling,sakaguchi2017homochiral,mutlu2021transfer,jacobberger2015direct,way2018seed,shekhirev2017interfacial} Such electrodes are fabricated using standard lithography techniques that may lead to contamination and/or the introduction of damage to the GNRs in the fabrication process. An appealing alternative is graphene. Being naturally atomically flat, it is optimally suited for ``GNR-last'' fabrication approaches. Graphene electrodes are either defined using electron beam lithography (EBL) nanogaps\citep{braun2021optimized,chen2016synthesis,candini2017high} (Fig.~\ref{fig:intro}b , middle) or formed using the electrical breakdown (EB) method that results in ultra-narrow nanogaps in the range of 1-5nm\citep{el2020controlled,sun2020massive,martini2019structure,fantuzzi2016fabrication} (Fig.~\ref{fig:intro}b, right). However, for both metallic and graphene electrodes, it is still challenging to contact an individual GNR because of their intrinsic width and lateral separation (Fig.~\ref{fig:intro}a), both typically in the order of $\approx$1-2~nm. These dimensions are well below the capabilities of state-of-the-art electron beam lithography (Fig.~\ref{fig:intro}a, gray dash lines)\citep{qiu2017scaling,chen2021sub}. Alternatively, the inter-ribbon separation between the GNRs could be increased, but this is usually achieved by reducing the amount of precursor molecules on the growth substrate and comes at the cost of shorter GNRs.\citep{lin2022scaling} Nevertheless, individual GNRs have been contacted previously using graphene-based breakdown gaps.\citep{ElAbbassi2020ControlledQuantumDot} However, this method yields ill-defined electrode geometries and it only works for very short GNRs that are comparable to the electrode separation ($\approx$5~nm), as for longer ones, the probability of having multiple GNRs bridging is substantially increased. For longer GNRs, for example such as those that can host a spin chain\citep{mishra2021observation} or topologically protected states\citep{Groning2018,Rizzo2018TopologicalBandEngineering}, it may be desirable to have the functional part located in-between the electrodes, rather than on top of the electrodes. Therefore, for longer GNRs, alternative contacting methods are required.\\

In this work, we demonstrate the contacting of single, long GNRs in a multi-gate transistor geometry using one-dimensional (1D) single-walled carbon nanotubes (SWNTs) as ultimately-scaled electrodes (Fig.~\ref{fig:intro}c) with diameters as small as $\sim$1 nm. Our approach relies on the self-aligned nature of both the SWNT and GNR growth on their respective growth substrate. The successful assembly of the SWNT-GNR-SWNT devices is verified using spectroscopy of the molecular levels performed at cryogenic temperature, with the observation of several features that are characteristic of transport through an individual GNR, such as Coulomb blockade, the presence of vibrational modes in the single electron tunneling regime, and Franck-Condon blockade. Moreover, the multiple gates allow for individually tuning of the conductivity of the GNRs and the SWNT electrodes, as well as identify the origin of the different states observed in the spectroscopic measurements.

\section{Concept and design}

\begin{figure}
\begin{center}\includegraphics[width=\textwidth]{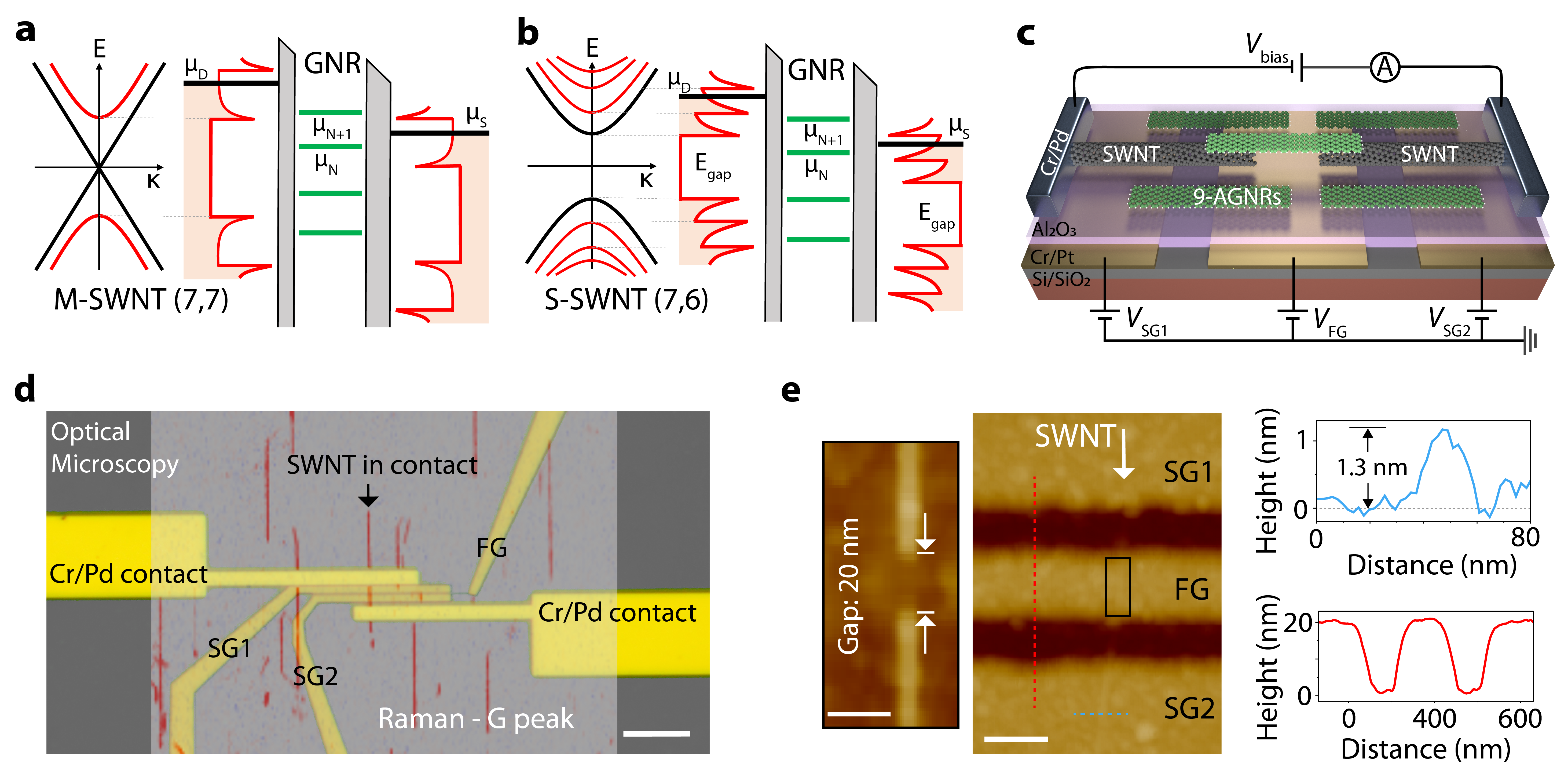}
          \end{center}
\caption{ \textbf{Multi-gate 9-AGNR transistors with SWNT electrodes.} 
(\textbf{a}) Left: The electronic dispersion relation of a representative M-SWNT (7,7) with subbands and zero bandgaps. Right: Illustration of the discrete energy levels of the GNR and sharp density of states (DOS) in the two SWNT electrodes. The sharp DOS peaks exhibit van Hove singularities which are associated with the subbands of SWNT. Note that we indicate the barriers between the left SWNT electrode and GNR, and between the right SWNT electrode and GNR. Asymmetrical barriers are illustrated in practice.
(\textbf{b})Similar illustration as (\textbf{a}) but for a representative S-SWNT (7,6) with the same diameter as the SWNT in (\textbf{a}). It has relatively more dense van Hove singularities in DOS and a finite bandgap. The electronic dispersion and DOS for SWNTs (7,7) and (7,6) are adapted from REF\citep{odom2000structure}. Note that the chiralities of the SWNTs used for this work were not determined.
 (\textbf{c}) Schematic illustration of the device, including the measurement circuit.  
 (\textbf{d}) Optical image of a device with an overlay of the G-peak Raman intensity map colored in red (532~nm laser, 1~mW power and 1~s integration time). Scale bar: 10~µm.  
(\textbf{e}) Topographic AFM image showing the SWNT electrodes and gate layout of the device. Left: High-resolution AFM characterization of a representative SWNT nanogap ($\sim$ 20~nm) defined by e-beam lithography. Scale bar: 20~nm. Middle: Topography profile across the SWNT (blue solid line) and gate electrodes (red solid line) with a 30 nm ALD Al$_{2}$O$_{3}$ layer on top. Scale bar: 100~nm. Right: The profiles (blue and red solid line) are taken along the blue and red dot lines in the corresponding AFM image, respectively. 
  }	
    \label{fig:device}
\end{figure}

The studied devices (Fig~\ref{fig:device}) consist of a pair of SWNT electrodes separated by 15-25~nm. Below the nanogap, a 100~nm wide Cr/Pt finger gate (FG) is fine-patterned alongside the two side gates (SG1 and SG2). The multiple gates are required for controlling the density of states of the SWNT leads. Due to quantum confinement of the charge carriers as a result of the one-dimensional nature of the SWNTs, sharp peak-like van Hove singularities appear at the onset of each subband\citep{mintmire1998universal,odom2000structure}. In addition, SWNTs come in two types, metallic SWNTs (M-SWNTs) and semiconducting SWNTs (S-SWNTs). While metallic SWNTs exhibit a flat and non-zero density of states around the Fermi energy, their semiconducting counterparts have a sizable bandgap. Fig~\ref{fig:device}a and Fig~\ref{fig:device}b illustrate the band diagrams of the SWNT-GNR-SWNT junctions with the discrete energy levels of the GNR and the van Hove singularities in the density of states of the M-SWNT and S-SWNT leads, respectively. The multiple gates are separated from the junction by a 30~nm thick Al$_{2}$O$_{3}$ layer. A film of GNRs is then transferred on top of the device substrate. Fig~\ref{fig:device}c shows a schematic of the device. A detailed description of the materials and the fabrication process can be found in Materials and Methods, and in Section 1 of the Supplementary Information.\\

Fig~\ref{fig:device}d presents an optical image of a representative device, with three gates and source/drain contacts. As red overlay, we present the Raman intensity map of the G-peak, highlighting the presence of the uniaxially aligned SWNT array, with a single SWNT bridging the metallic source/drain contacts. Fig~\ref{fig:device}e shows an atomic force micrograph (AFM) of the device focusing on the gate structure and the SWNT electrodes. A high-resolution AFM image of the SWNT nanogap is presented as inset, revealing a gap size of $\sim$ 20~nm and a SWNT diameter of 1.3~nm. More characterizations of the SWNT diameters can be found in section 1.1 of the Supplementary Information. As the SWNT diameter is of similar size as the GNR width, we anticipate that one, or at most two GNRs can make contact to a pair of SWNT electrodes.

In total, 8 chips were characterized, 5 with the multi-gate architecture, and 3 with a single back-gate, for a total of $\sim$ 2500 devices. This number only includes devices on which GNRs were transferred, not those present on the chips but located outside the area covered by the GNR films. Among them, $\sim$ 600 SWNT transistors were functional at room temperature, as assessed by electrical characterization, before the nanogap formation. After the nanogap formation, 360 devices showed clearly separated SWNT electrodes, with currents lower than 10 pA at 1~V. After the GNR transfer, 41 of those devices showed gate-modulated current. GNR films are known to conduct at room temperature\cite{Richter2020ChargeTransportMechanism} and one therefore cannot attribute these devices to nanogaps containing only individual GNRs. However, as film transport is temperature activated, it is easily suppressed by cooling down the sample to cryogenic temperatures. At temperatures below 9~K, 12 devices showed quantum dot (QD) behavior. This corresponds to a yield of 3.3\% when considering only the number of nanogaps that were well formed before the GNR transfer. The details of the electrical characterization are provided in the Materials and Methods, and in Section 1.3 of the Supplementary Information. \\

In the following, we discuss QD devices based on M-SWNT leads (Devices: D3 and D6) and S-SWNT leads (D7), all obtained using the multi-gate architecture and characterized at a base temperature of 255 mK using a $^3$He system. In the Supplementary Information, we present additional devices, either based on the global back-gate (D1 in Section 2) or multi-gate (D4, D5 and D8 in Section 3) architecture. 

\section{Multi-gate devices with M-SWNT leads}

\begin{figure}
    \begin{center}\includegraphics[width=\textwidth]{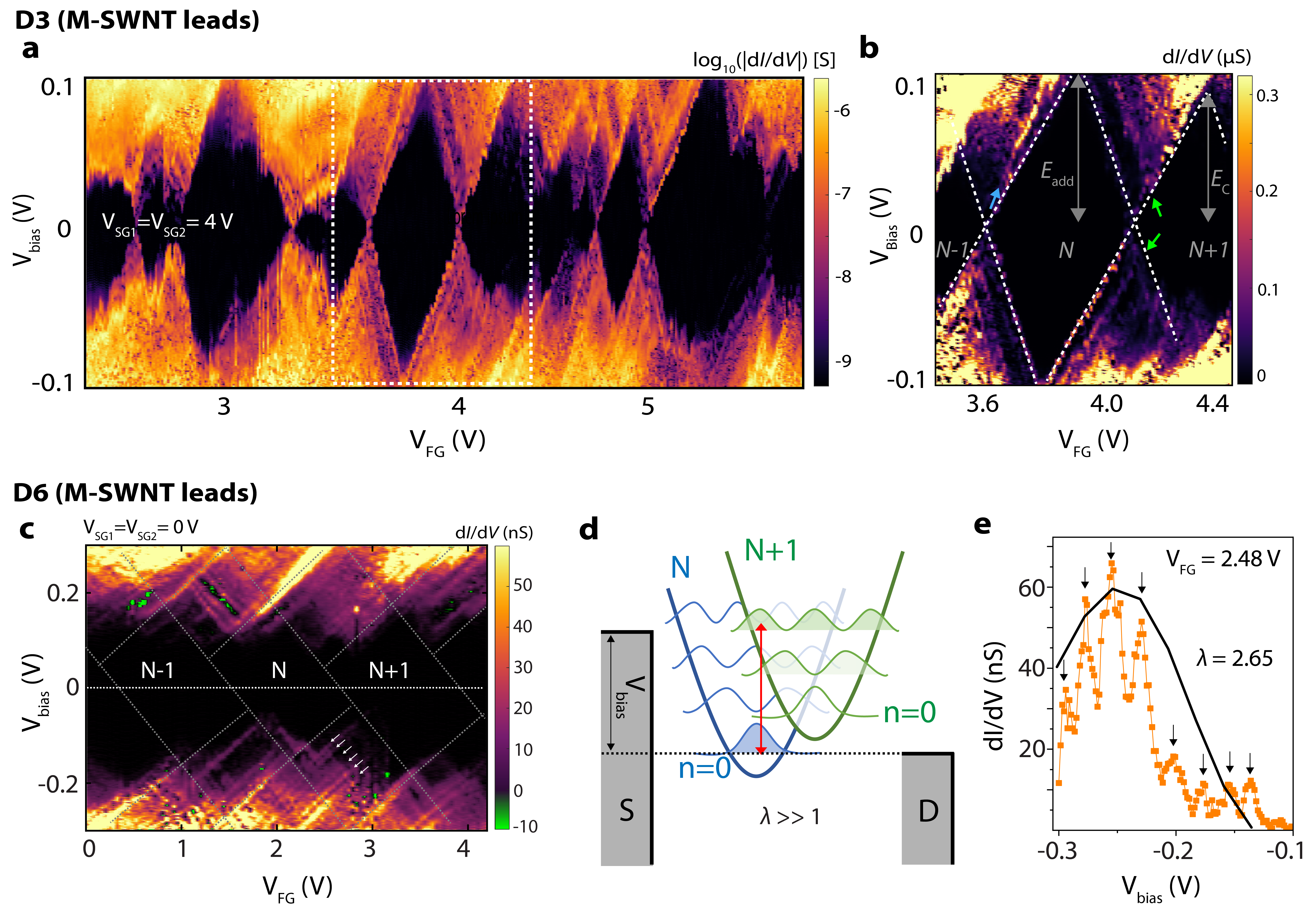}\end{center}
\caption{\textbf{Electron Transport in 9-AGNR transistors (D3 and D6) with M-SWNT leads.} 
    (\textbf{a, b}) Single electron charging behavior in D3. (\textbf{a}) Colour scale differential conductance versus V$_{\textnormal{FG}}$ and V$_{\textnormal{Bias}}$ at fixed V$_{\textnormal{SG1}}$ = V$_{\textnormal{SG2}}$ =4 V, showing single electron charging behaviour. (\textbf{b}) Close-up of the box in (\textbf{a}, white dot-line) highlighting the excited states (green arrows) and lead states (blue arrow). 
    (\textbf{c-e}) Franck-Condon blockade in D6. (\textbf{c}) Colour scale differential conductance versus V$_{\textnormal{FG}}$ and V$_{\textnormal{Bias}}$ for V$_{\textnormal{SG1}}$ = V$_{\textnormal{SG2}}$ = 0~V. Low-bias conductance is suppressed and the Coulomb blockade cannot be lifted by V$_{\textnormal{FG}}$. Periodic excitations (white arrows) appear within the conductive regime at positive and negative bias. NDC appears in some regions (green color).  
    (\textbf{d}) Schematic representation of the Franck-Condon model for strong electron-phonon coupling $\lambda$, with N (blue curve) and N+1 (green curve) electrons in the QD. The tunneling electron shifts the equilibrium coordinate of the phonon harmonic oscillator by an amount proportional to $\lambda$, thereby exponentially suppressing the transition between the vibrational ground states of the $N$ and $N+1$ charge states.  (\textbf{e}) Differential conductance measured for V$_{\textnormal{FG}}$ = 2.48~V and V$_{\textnormal{SG1}}$ = V$_{\textnormal{SG2}}$ = 0~V. Representative fit of the maxima with the Franck–Condon progression (eq.~\ref{eq:FC}) enable us to extract the coupling $\lambda$  = 2.65. 
  }
    \label{fig:MSWNT}
\end{figure}

Fig~\ref{fig:MSWNT}a-b present the transport data for D3 with a pair of M-SWNT leads. Fig~\ref{fig:MSWNT}a shows the differential conductance (dI/dV) as a function of V$_{\textnormal{FG}}$ and V$_{\textnormal{Bias}}$ (so-called stability diagram) for fixed side gate voltages of V$_{\textnormal{SG1}}$ = V$_{\textnormal{SG2}}$ = 4~V. For the given gate voltage range, several Coulomb diamonds are observed with strong variations in the addition energies, ranging from 33 to 110~meV. A close-up of the boxed region in Fig~\ref{fig:MSWNT}a (see Fig~\ref{fig:MSWNT}b) shows a well-resolved single-electron tunneling (SET) regime with multiple resonances that run parallel to the edge of the diamond, as marked by the green arrows. For the SET regime around a gate voltage of 4~V, the excited states at positive and negative bias are located at 25~mV and -23~mV (green arrows), respectively. We attribute these resonances to the presence of vibrational modes in the 9-AGNRs, which will be discussed further later. 
To confirm the conductive nature of the electrodes due to the absence of a bandgap, we measured the stability diagram in a different transport regime by applying side gate voltages of V$_{\textnormal{SG1}}$ = V$_{\textnormal{SG2}}$ = 0~V, yielding qualitatively similar results (see section 3.1 of Supplementary Information). \\

In addition to the excited states, we observe additional resonances that we attribute to the modulation of the DOS in the SWNT leads\citep{thomas2021spectroscopy}, as highlighted by the blue arrow in Fig~\ref{fig:MSWNT}b. These states can be distinguished by their slope $\Delta V_{Bias}$ / $\Delta V_{FG}$ in the stability diagram, which is different from the slopes of the edge of the SET regime. The origin of such slope difference is the different gate coupling of the FG to the GNRs and SWNT leads. Additionally, we observe several Coulomb diamonds in Fig~\ref{fig:MSWNT}a that do not have a crossing point at each of their sides, e.g., at V$_{\textnormal{FG}}$ of $\sim$ 3~V, $\sim$ 4.4~V, and $\sim$ 5.4~V. This may be due to the mixing of the lead states with QD states\citep{guttinger2009electron}. In Section 3.2 of Supplementary Information, two additional devices (D4 and D5) with M-SWNT leads are shown, with qualitatively similar behavior. \\

Fig~\ref{fig:MSWNT}c-e show the transport data for D6 with a pair of M-SWNT leads. Although D6 has a similar fabrication process as D3-D5, richer physics is observed. Fig~\ref{fig:MSWNT}c shows a stability diagram for fixed V$_{\textnormal{SG1}}$ = V$_{\textnormal{SG2}}$ = 0 V. In the given V$_{\textnormal{FG}}$ ranges, Coulomb diamonds are observed, possessing several striking features. First, quasi-periodic lines (white arrows) running parallel to the edges of the Coulomb diamonds are observed when V$_{\textnormal{Bias}}$ $>$ +60 meV and V$_{\textnormal{Bias}}$ $<$ -60 meV. The energy spacing between these excited states is $\Delta E_{}$ = 29~meV on average. Second, conductance is highly suppressed at low bias regime between $\sim$ +60 meV and $\sim$ -60 meV present in all of the probed diamonds. Third, in some regions, negative differential conductance (NDC) appears in between resonances, as indicated by the green color. We attribute these three features to phonon-assisted tunneling transport, enabled by a strong electron-phonon coupling in our GNR junction, as discussed in more details below.\\

Similar quasi-periodic resonances have been previously attributed to the excitation of vibrational modes, as has been observed in single-molecule transistors\citep{park2000nanomechanical,lau2016redox,burzuri2014franck,pasupathy2005vibration} and suspended SWNTs\citep{sapmaz2006tunneling,leturcq2009franck}. The zero-bias conductance suppression may originate from the Franck-Condon blockade effect\citep{leturcq2009franck,lau2016redox,burzuri2014franck,weig2004single}, which occurs in case of strong electron-phonon coupling ($\lambda$ $\gg$ 1), as described in Fig~\ref{fig:MSWNT}d. Here, sequential electron tunneling is strongly suppressed due to the exponentially small overlap of the harmonic oscillator wavefunctions of the different charge states, and charge transport can only occur when the bias is large enough to overcome the phononic energy difference by exciting phonons. To extract the electron-phonon coupling $\lambda$, we study the dI/dV versus V$_{\textnormal{Bias}}$ at a fixed V$_{\textnormal{FG}}$ = 2.48~V (see Fig~\ref{fig:MSWNT}e). By fitting the maxima of dI/dV with the Franck-Condon model (see Methods section for more details), $\lambda$ is determined to be 2.65. The average $\lambda$ obtained from another four dI/dV traces at different transport regime is 2.66 ± 0.09 (see Section 3.3 of Supplementary Information). Overall, the $\lambda$ is symmetric with respect to the bias polarity and independent of the charge state, which is consistent with a phononic origin. Another interesting feature of the data is the appearance of the NDC in Fig~\ref{fig:MSWNT}c. Such NDC regions have previously been associated with electron-phonon interactions, according to theoretical\citep{nowack2005vibration,mccarthy2003incoherent,koch2005franck} and experimental\citep{sapmaz2006tunneling,leturcq2009franck} studies. Importantly, the three above-mentioned transport features require a strong electron-phonon coupling, as well as the presence of a single QD in the junction area.  \\

To determine the position of the QD along the SWNT-GNR-SWNT channel, we have measured the current as a function of the multiple gates and extracted the relative gate couplings $\alpha_{\textnormal{FG}}$ : $\alpha_{\textnormal{SG1}}$ : $\alpha_{\textnormal{SG2}}$  = 1 : 0.81 : 0.29. More detailed about the analysis of the gate coupling can be found in Section 3.4 of Supplementary Information. Based on these and the position of the gates with respect to the SWNT-GNR-SWNT channel, we conclude that the QD is formed in the GNR, rather than the in SWNT leads.

\section*{Electron and phonon properties of 9-AGNRs}

\begin{figure}
\begin{center}
	  \includegraphics[width=0.59\textwidth]{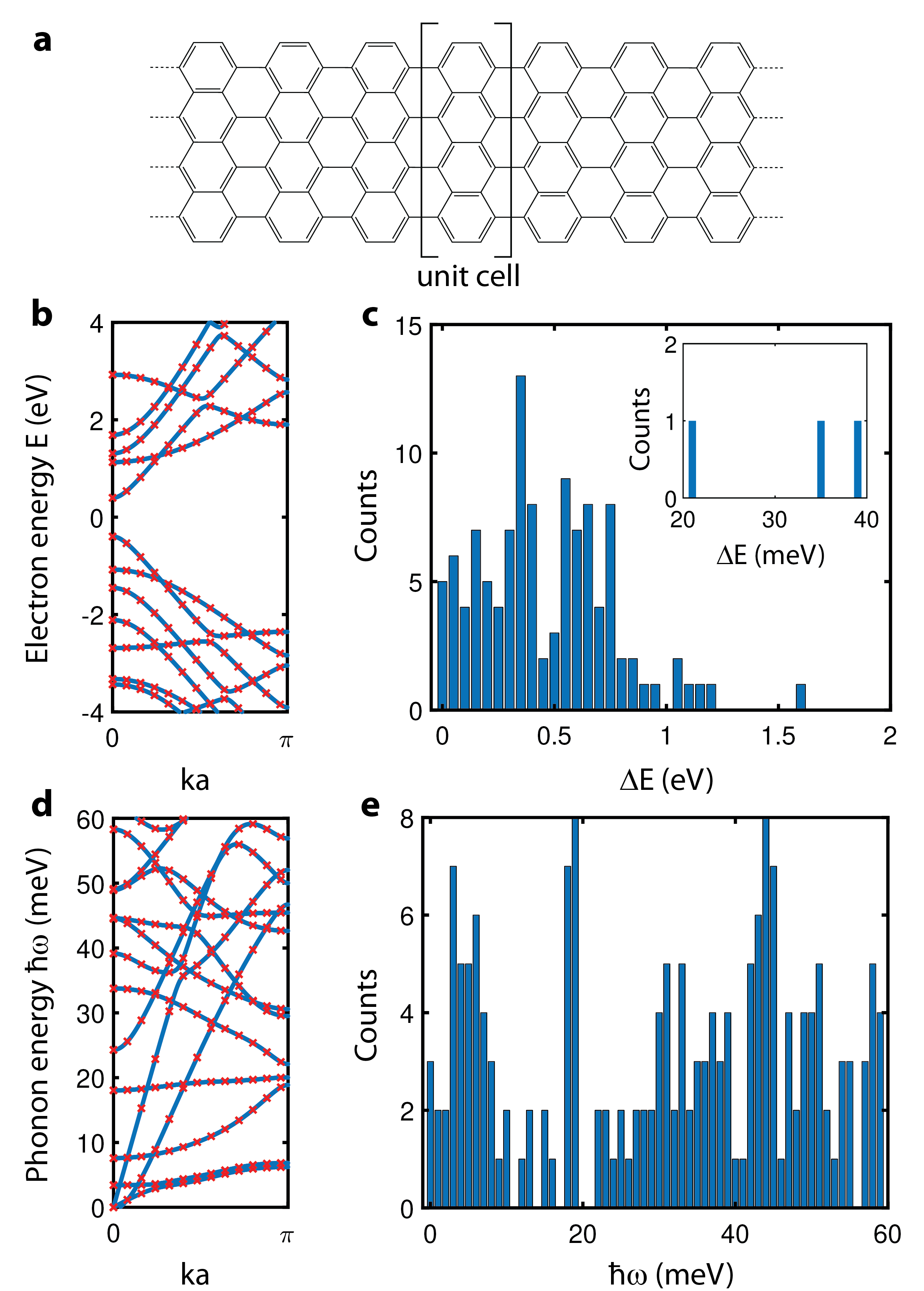} 
\end{center}
\caption{\textbf{Electron and phonon properties of 9-AGNRs.} 
(\textbf{a}) Molecular structure of a 9-AGNR.
(\textbf{b}) Electron band structure of 9-AGNR. The energy scale is within $\pm$ 4 eV.
(\textbf{c}) Histogram of electron energy level spacing of 9-AGNR determined by the energy differences of neighboring bands at various $k a$ values (red cross in \textbf{b}). The inset present a zoom-in in the region 20-40 meV.
(\textbf{d}) Phonon band structure of 9-AGNR. The energy scale is within 0 - 60 meV.
(\textbf{e}) Histogram of phonon energy of 9-AGNR determined by the energies of different bands at various $k a$ values (red cross in \textbf{d}).
    }
    \label{fig:theory}
\end{figure}

To rationalize the charge-transport measurements, and in particular to identify the origin of the resonances observed in the SET regime of devices D3 and D6, we turn to quantum chemistry calculations. We initially compute the electron and phonon band structures by performing periodic density functional theory calculations (GGA-PBE) of a 9-AGNR unit cell (see Materials and Methods - Computational Methods for more details). To account for the quantum confinement effect in a finite-length 9-AGNR of 60~nm, we discretize the two band structures and obtain the corresponding energy levels\cite{Sadeghi2018TheoryElectronPhonon}. Fig.~\ref{fig:theory}a shows a schematic illustration of the 9-AGNR, alongside the calculated electronic band structure in Fig.~\ref{fig:theory}b. The plot displays a semiconducting behavior with a bandgap of $\sim$ 796~meV, in agreement with the previous electronic band structure calculations\cite{Son2006EnergyGapsGraphene}. The red crosses on the band structure graphs correspond to the discretized energy levels of the 9-AGNR. From the energy levels spectrum, we build a histogram of the energy differences of adjacent energy levels for the selected \textit{ka} values, shown in Fig.~\ref{fig:SSWNT}c.\\

The plot shows that for all $k a$ values combined, most of the energy spacings are on the order of hundreds of mili-electonvolts up to electonvolts, with hardly any counts in the tens of millielectron volt range, and no counts between 22 and 34~meV (see inset). Fig.~\ref{fig:theory}d shows the calculated phonon band structure for energies up to 60~meV, including the discretized values, with in Fig.~\ref{fig:theory}e a histogram of all the vibrational modes. The histogram possesses tens of modes in the 20-30~meV range, which is comparable to the experimentally observed values of 23-25~meV for device D3, and more than one order of magnitude smaller than the typical level spacings computed for electrons\cite{sapmaz2006tunneling} (Fig.~\ref{fig:theory}c). From the absence of electronic energy spacings in the 22 to 34~meV range (Fig.~\ref{fig:theory}c) and the dense population of vibrational modes in the same range, we attribute the excited states in device D3 to vibrational modes. This observation is also in line with the observed equidistant resonances and low-bias gap in device D6 being caused by Franck-Condon blockade. Indeed, the observed equidistant energy spacing $\Delta$ E ($\sim$ 29 meV) is consistent with the low-energy regime (20 - 40 meV) where many vibrational modes exist.

\section{Multi-gate devices with S-SWNT leads.}

\begin{figure}
    \begin{center}\includegraphics[width=\textwidth]{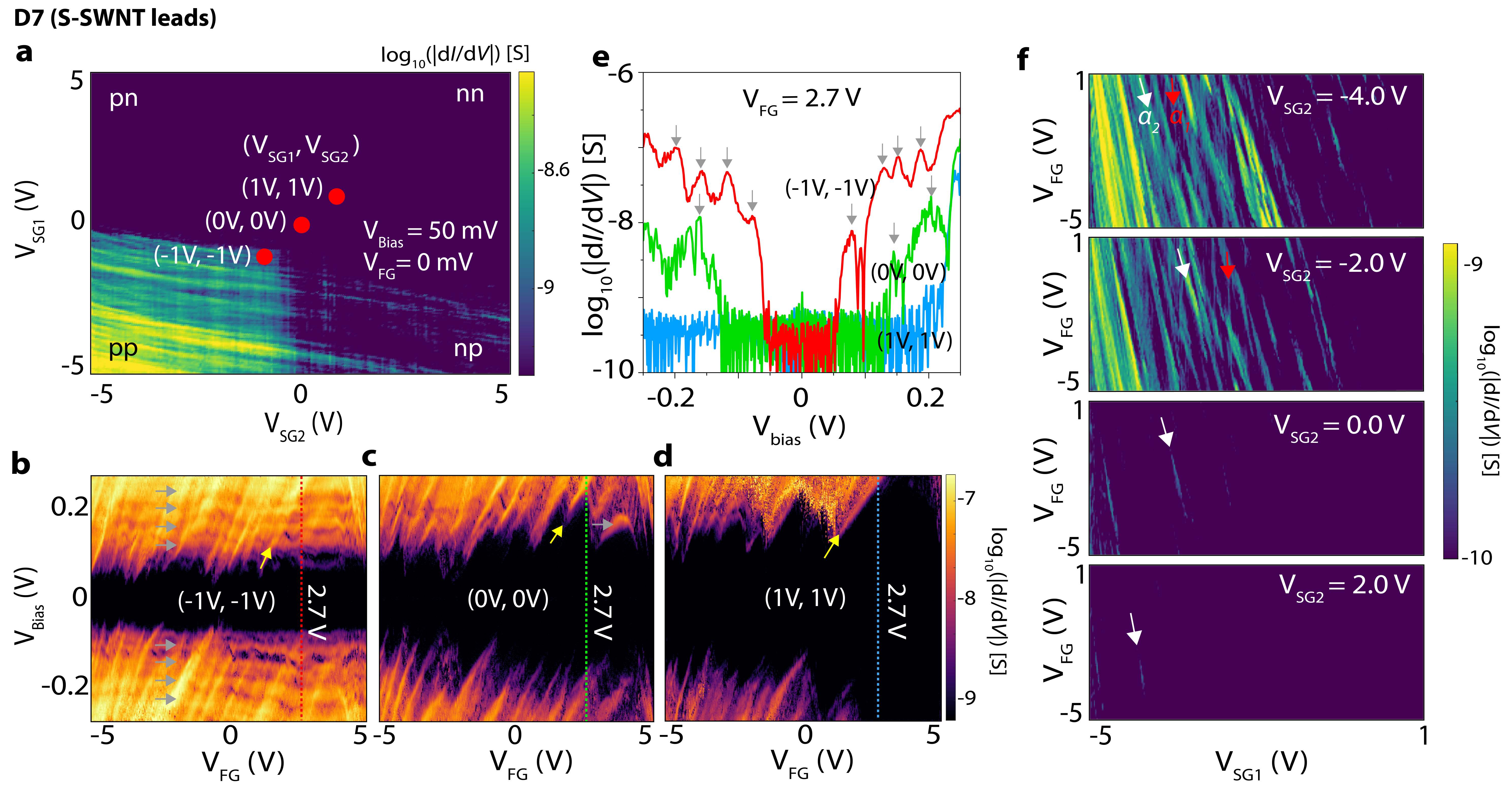}\end{center}
\caption{\textbf{Electron Transport in 9-AGNR transistors (D7) with S-SWNT leads.}
(\textbf{a}) Differential conductance as a function of V$_{\textnormal{SG1}}$ and V$_{\textnormal{SG2}}$ at fixed V$_{\textnormal{Bias}}$ = 50~mV, and V$_{\textnormal{FG}}$ = 0~V. The three red dots mark the positions of (V$_{\textnormal{SG1}}$, V$_{\textnormal{SG2}}$) settings for the measurements in b-d.
(\textbf{b-d}) Colour scale differential conductance versus V$_{\textnormal{FG}}$ and V$_{\textnormal{Bias}}$ for different (V$_{\textnormal{SG1}}$, V$_{\textnormal{SG2}}$) settings: (\textbf{b}), (-1~V, -1~V); (\textbf{c}), (0~V, 0~V); (\textbf{d}), (1~V, 1~V). 
(\textbf{e}) Differential conductance as a function of V$_{\textnormal{Bias}}$ cut along V$_{\textnormal{FG}}$ = 2.7~V indicated by the red dot line (-1~V, -1~V) in (\textbf{b}), the green dot line (0~V, 0~V) in (\textbf{c}) and the blue dot line (1~V, 1~V) in (\textbf{d}), respectively. 
(\textbf{f}) Differential conductance as a function of V$_{\textnormal{FG}}$ and V$_{\textnormal{SG1}}$ for different V$_{\textnormal{SG2}}$ of -4~V, -2~V, 0~V and 2~V. Several resonance lines are observed with two slopes, as highlighted by a red arrow and a white arrow.
    }
    \label{fig:SSWNT}
\end{figure}

We now turn to the use of S-SWNTs, for which the the transport measurements on device D7 are shown in Fig.~\ref{fig:SSWNT}. Fig.~\ref{fig:SSWNT}a displays the dI/dV as a function of V$_{\textnormal{SG1}}$ and V$_{\textnormal{SG2}}$ for a fixed V$_{\textnormal{Bias}}$=50 mV and V$_{\textnormal{FG}}$ = 0~V. This conductance map shows that the leads are conductive when a negative voltage (p side) is applied to either of the side gate, while transport is suppressed for positive voltages (n side). As a result, depending on the combination of SG voltages, the device can be tuned in four different regimes: p-n, n-n, n-p or p-p. \\

We performed stability diagrams (V$_{\textnormal{Bias}}$ versus V$_{\textnormal{FG}}$) for the three combinations of (V$_{\textnormal{SG1}}$, V$_{\textnormal{SG2}}$), highlighted with red dots, gradually transitioning from the p-p to the n-n regime. The differential conductance maps are presented in Fig.~\ref{fig:SSWNT}b-d. Several characteristic features can be observed. First, there is a region at low bias for which the current is blocked. In the high-bias regime, where current is flowing, two typical resonances are observed: resonances that are highly affected by the finger gate voltage (in the following be referred to as gate-dependent resonance, with one representative highlighted by the yellow arrow) and resonances that are mostly unaffected by the finger gate (gate-independent resonance), resulting in predominantly horizontal lines in the stability diagram (highlighted by the grey arrow). The two different features are exemplary illustrated in Fig.~\ref{fig:SSWNT}b. Here, a low-bias gap is present up to voltages of about 50~mV while the high-bias regime exhibits both gate-dependent and gate-independent resonances. We notice that the gate-dependent resonances have predominantly positive slopes, which we attribute to asymmetric tunneling rates between the GNR QD and the leads. Similar resonances are also observed in another device (D8), as shown in Section 3.5 of Supplementary Information. For V$_{\textnormal{SG1}}$ = V$_{\textnormal{SG2}}$ = 0~V, the low-bias gap increases to $\sim$ 80-110~mV. Moreover, most of the gate-independent resonances have disappeared, while the gate-dependent ones remain mostly unaffected by the side gate voltages, although a slight decrease in intensity is observed. For V$_{\textnormal{SG1}}$ = V$_{\textnormal{SG2}}$ = 1~V, the low-bias gap is further increased and the gate-independent resonances remain mostly absent, while the gate dependent ones, although still present, have reduced in intensity. To better visualize the effect of the side gates, Fig~\ref{fig:SSWNT}e presents the dI/dV as a function of V$_{\textnormal{Bias}}$ at a fixed V$_{\textnormal{FG}}$ = 2.7~V and for different SG settings (V$_{\textnormal{SG1}}$, V$_{\textnormal{SG2}}$) settings, \textit{i.e.}, linecuts from Fig.~\ref{fig:SSWNT}b-d obtained at the positions marked by the dashed lines. Here, the widening of the low-bias gap with increasing side gate voltages is clearly visible, as well as the multiple resonances for V$_{\textnormal{SG1}}$ = V$_{\textnormal{SG2}}$ = -1~V. To investigate further the two types of resonances observed in the stability diagram, we measured the dI/dV while sweeping the V$_{\textnormal{FG}}$ and the V$_{\textnormal{SG1}}$ for different fixed V$_{\textnormal{SG2}}$ (see Fig.~\ref{fig:SSWNT}f). This measurement allows us to determine the coupling of each of the resonances to the different gates. For V$_{\textnormal{SG2}}$ = -4~V, a range of resonances is observed, primarily moving downward with increasing voltage on SG2. One example of such resonance is highlighted by the white arrows. Careful inspection also reveals additional narrow and more vertical resonances, of which one is highlighted by the red arrows. The two lines have different slopes, representing the relative gate coupling between SG1 and FG: $\alpha_{\textnormal{1}}$ $\approx$ 52.5 and $\alpha_{\textnormal{2}}$ $\approx$ 6.9. For increasing V$_{\textnormal{SG2}}$, the resonances gradually fade out until they mostly disappear. \\

Based on the multiple gate-dependent measurements we have performed, we attribute the finger gate dependent resonances to states associated with the discrete energy levels of the GNR QD, as the finger gate is expected to couple more strongly to the GNR QD due to its close proximity. Conversely, the resonances that are largely unaffected by the finger gate originate from modulations in the DOS of the leads. These lead states could originate either from the pristine SWNT or from localized states due the presence of defects and/or other local barriers\cite{biercuk2004locally}. The position of the lead states is tunable with by the SGs, which, in addition, also have the two following effects: first, the conductance through the S-SWNT is being suppressed as the SWNT is being switched off. Second, when the resistance of the electrodes becomes comparable to that of the QD, the systems acts as a voltage divider, with part of the bias voltage dropping across the electrodes themselves. As a result, the effective voltage across the QD is reduced and the low-bias gap increases, as observed in Fig.~\ref{fig:SSWNT}b-d.\\

Overall, by comparing the metallic and semiconducting SWNT as electrode material, we find that metallic SWNT have a clear advantage of over semiconducting ones. First, no bandgap is present and the contacts can therefore not be switched off. In addition, devices with metallic SWNT electrodes possess fewer signatures of the electrodes themselves in charge transport measurements. 

\section{Conclusion}
 
In conclusion, we demonstrated the successful contacting of individual GNRs using pairs of SWNT leads as ultimately-scaled electrodes with diameters as small as $\sim$ 1 nm. The devices present behavior that is characteristic for charge transport through a single quantum dot, such as Coulomb blockade with excited states of vibrational origin and Franck-Condon blockade. Our findings are supported by DFT calculations that highlighted the importance of vibrational modes. 

Contacting single, long, GNRs using SWNT electrodes embedded in a multi-gate architecture represents a major step forward in the exploitation of their highly-tunable physical properties in electronic and spintronic devices. This is in particular relevant for GNRs that host physical phenomena based on long-range effects, such as spin-chains\cite{mishra2021observation} or the creation of topological bands due to the periodic placement of edge-extension along the GNR backbone\citep{Groning2018,Rizzo2018TopologicalBandEngineering}. These effects offer great promises for a range of quantum technologies such as quantum computing, quantum communication, and energy conversion.

\section*{Acknowledgements}
M.C. acknowledge funding by the EC H2020 FET Open project no. 767187 (QuIET) and by the Swiss National Science Foundation under the Sinergia grant no. 189924 (Hydronics). M.L.P. acknowledges funding by the Swiss National Science Foundation (SNSF) under the Spark project no. 196795 and the Eccellenza Professorial Fellowship no. PCEFP2\textunderscore203663. This work was supported by the Swiss State Secretariat for Education, Research and Innovation (SERI) under contract number MB22.00076. H.S. acknowledges the UKRI for Future Leaders Fellowship number MR/S015329/2. S.S acknowledges the Leverhulme Trust for Early Career Fellowship no. ECF-2018-375. G.B.B, P.R and R.F acknowledge funding by the Swiss National Science Foundation under grant no. 200020-182015, the European Union Horizon 2020 research and innovation program under grant agreement no. 881603 (GrapheneFlagship Core 3), and the Office of Naval Research BRC Program under the grant N00014-18-1-2708. We also greatly appreciate the financial support from the Werner Siemens Foundation (CarboQuant). We thank Michael Stiefel and the Cleanroom Operations Team of the Binnig and Rohrer Nanotechnology Center (BRNC) for their help and support. The authors thank Christian Schönenberger and Andreas Baumgartner for fruitful discussions.

\section*{Author contributions}
Jian Z., M.L.P. and M.C. conceived and designed the experiments. K.M. provided the GNR precursor molecules. G.B.B. under the supervision of P.R. and R.F., performed the on-surface synthesis and substrate transfer of GNRs. L.Q. under the supervision of Jin Z. performed the growth of aligned SWNTs and SWNT transfer. Jian Z., P.C. and L.Q. fabricated the devices. Jian Z. and M.L.P. performed the electrical measurements. Jian Z. performed the Raman and AFM measurements. Jian Z. and M.L.P. analyzed the data. A.D., S.S and H.S. provided the theory and performed the theoretical calculations. Jian Z., M.L.P., and M.C. discussed the figures and wrote the manuscript. All authors discussed the results and their implications and commented on the manuscript.

\section*{Competing interests}
The authors declare that they have no competing interests.

\section*{Data and materials availability}
All data needed to evaluate the conclusions in the paper are present in the paper and/or the Supplementary Materials. Additional data related to this paper may be requested from the authors.

\section*{Materials and Methods}
\subsection*{Fabrication and characterization}
The device is fabricated as follows. First, a 100~nm wide finger gate (FG) is fine-patterned alongside the two side gates (SG1 and SG2). These gates consist of 5/15~nm of Cr/Pt and are covered by 30~nm Al$_{2}$O$_{3}$ layer deposited using atomic layer deposition (ALD) acting as the gate dielectric. A large-area quartz crystal (4×6~$mm^2$) is used to synthesize a uniaxially aligned array of SWNTs, which is transferred on top of the aluminum oxide of the device chip using a wet-transfer method. The SWNT transistors are then fabricated by depositing a periodic array of metallic pads (3/50~nm of Cr/Pd) to contact the SWNTs. The overall channel length between the two metal electrodes is 2.5~µm. As the width of the metal contacts (10~µm) is comparable to the average separation distance between in the SWNTs, we assume that most of the as-fabricated transistors contain only a single SWNT. Then, nanogaps of 15-25~nm are formed in the SWNTs using an optimized electron-beam lithography process in combination with reactive ion etching. Here, the electrode separation was set to be 15-25~nm, large enough to eliminate direct tunneling contributions between the electrodes, but much smaller than the average length of the 9-AGNRs.  Finally, a dense array of uniaxially aligned 9-atom wide armchair GNRs (9-AGNRs) is grown on a Au(788) substrate and transferred on top of the device\cite{Overbeck2019OptimizedSubstratesMeasurement}. The integrity of the 9-AGNRs and their alignment with respect to the source-drain axis were confirmed using polarization-dependent Raman spectroscopy. In Section 1.1 and 1.2 of the Supplementary Information, a more detailed description of the fabrication process is given. 

Note that devices with both the global back-gate (BG) architecture and multi-gate architecture were fabricated in this work. The electrical characterizations were performed at four different steps during the device fabrications using DC measurement techniques: first, After patterning electrode arrays in the transferred SWNT area, the gate-modulated electrical conductivity for each defined channel was measured. The purpose here was to screen the SWNT transistors and to determine the electrical properties of SWNTs at each transistor. Second, after forming the nanogaps on SWNT by EBL, the SWNT transistors were electrically characterized to ensure a clear separation between the electrodes. Devices with currents $\textgreater$ 20 pA at V$_{\textnormal{Bias}}$ = 1 V were excluded from further characterization. Third, after the transfer of GNRs on SWNT electrodes, electrical measurements were performed to find devices bridged by GNRs, which were selected for low-temperature measurements. The electrical measurements for steps 1-3 were performed at room temperature using an automatic probe station. Forth, the pre-selected GNR devices were measured at low-temperature under vacuum conditions ($<$10$^{-6}$~mbar).  The global BG devices (D1, D2) were measured in commercially available cryogenic probe station (Lake Shore Cryogenics, Model CRX-6.5K) at a base temperatures of 9~K and the multi-gate devices (D3-D8) were measured in commercially available Helium-3 refrigerator (Oxford Instruments, Model HelioxVL) at a base temperature of 255~mK. A data acquisition board (ADwin-Gold II, J\"ager Computergesteuerte Messtechnik GmbH) is employed to apply the bias and gate voltages and read the voltage output of the I–V converter (DDPCA-300, FEMTO Messtechnik GmbH).

All devices were measured in a two-terminal setup, where we applied a bias voltage and measured the current, from which the differential conductance was calculated by taking the numerical derivative.

\subsection*{SWNT growth, transfer, and nanogap formation}
The catalyst precursor was Fe(OH)$_{\textnormal{3}}$/ethanol solution with a concentration of 0.05 mol $L^{-1}$ and the growth substrate was ST-cut quartz (single side polished, miscut angle $\textless$ 0.5\degree, surface roughness $\textless$ 5 {\AA}). After cleaning, the quartz substrates were annealed at 900 ℃ in air for 8 h for better crystallization. Before growth, the catalyst precursor was spin-coated onto the substrates at a speed of 2500 rpm. Then, the quartz substrates with dispersed catalyst precursor were put into a 1-inch tube furnace and heated in air to 830 ℃. After the system was purged with 300 standard cubic centimeters per minute (sccm) argon for about 10 min, a flow of hydrogen (200 sccm) was introduced for 8 min to reduce the catalyst precursor to form Fe catalyst nanoparticles. Then, an extra argon flow (50 $\sim$ 150 sccm) through an ethanol bubbler was introduced for the growth of SWNT arrays for about 15 min. After the growth, the tube was cooled to room temperature with argon gas protection. The density of the grown SWNT arrays could be adjusted by the flow rate of ethanol.

The transfer of SWNT arrays from quartz substrate to the target substrate was conducted with the assistance of poly(methyl methacrylate) (PMMA). In detail, PMMA (MW= 950K) was spin-coated onto the SWNT arrays at a speed of 3000 rpm and was heated to 150 ℃ for about 15 min. The PMMA film, encapsulated SWNT arrays, was separated from the quartz substrate in KOH aqueous solution (1 mol $L^{-1}$, 70 ℃). Then the PMMA/SWNTs film was attached to the target substrate and cleaned by ultrapure water. After drying at 60 ℃ for about 3 h, most of the PMMA was removed by hot acetone. The residual PMMA was further removed by decomposing at 450℃ in argon and hydrogen atmosphere for 2 h.

To define SWNT nanogaps, a 60 nm thick CSAR resist (ARP 6200.04, Allresist GmbH) was spin-coated. Following the second electron beam exposure, the resist was developed using a suitable developer (AR 600-546, Allresist GmbH) at room temperature for 1 min followed by an IPA rinse. Reactive ion etching, RIE (15~sccm Ar, 30~sccm O$_{\textnormal{2}}$, 25~W, 18~mTorr) for 12~s was used to cut the SWNT segment within the CSAR gap. After RIE, the etching mask was removed by immersing in 1-Methyl-2-pyrrolidinone (NMP) (Sigma Aldrich) at room temperature for 10 min followed by 60~min at 80~℃, cooled down for 30~min, rinsed with IPA, and blown dry with N2. This approach yields clean and well-separated SWNT electrodes (15-25~nm nanogaps). Similar process has been used to make clean graphene nanogaps, as reported elsewhere\citep{braun2021optimized}.

\subsection*{Characterization of SWNT nanogaps}
The separation of SWNT electrodes (gap size) was assessed using scanning electron microscopy (SEM) (Helios 450, FEI), and high-resolution atomic force microscopy (AFM) (Icon, Bruker) was employed to determine the electrode separation. The AFM was equipped with a sharp cantilever (tip radius = 2~nm) (SSS-NCHR-20, Nanosensors) operated in soft-tapping mode. The gap size is typically between 15~nm and 25~nm.

\subsection*{On-surface synthesis of aligned 9-AGNRs and transfer to device substrate}
9-AGNRs were synthesized from 3',6'-diiodo-1,1':2',1"-terphenyl (DITP). \citep{DiGiovannantonio2018surfaceGrowthDynamics} Using a Au(788) single crystal (MaTeK, Germany) as growth substrate results in uniaxially aligned 9-AGNRs (GNRs grown along the narrow Au(111) terraces).\citep{Overbeck2019OptimizedSubstratesMeasurement} The Au(788) surface was cleaned in ultra-high vacuum by two sputtering/annealing cycles: 1~kV Ar$^{+}$ for 10~min followed by annealing at 420~$^\circ$C for 10~min. Next, the precursor monomer DITP was sublimed onto the Au(788) surface from a quartz crucible heated to 70~$^\circ$C, with the substrate held at room temperature. After deposition of about 60\%-70\% of one monolayer of DITP, the substrate was heated (0.5~K/s) to 200~$^\circ$C with a 10~min holding time to activate the polymerization reaction, followed by annealing at 400~$^\circ$C (0.5~K/s with a 10~min holding time) to form the GNRs via cyclodehydrogenation.
The average GNR length is between 40 and 45~nm.\citep{DiGiovannantonio2018surfaceGrowthDynamics} 9-AGNRs were transferred from their growth substrate to the silicon-based substrates with predefined SWNT electrodes by an electrochemical delamination method using PMMA as described previously.\citep{Overbeck2019OptimizedSubstratesMeasurement, senkovskiy2017making, Overbeck2019UniversalLengthdependent}

\subsection*{Determination of electron-phonon coupling using Franck-Condon principle}

From the Franck-Condon principle, the transition probability from the $N$ state to the $N + 1$ charge state is given by the FC factor\citep{koch2006theory}:

\begin{equation}
P_{\textnormal{m,0}} \propto (\frac{dI}{dV})_m^{max} \propto \frac{\lambda^{2m}}{m!} e^{-\lambda^2}   
\label{eq:FC} 
\end{equation}
where $m$ is the difference in phonon quantum numbers.

\subsection*{Computational methods}
The optimized geometry, ground state electron Hamiltonian and overlap matrix elements as well as phonon dynamical matrix of each structure studied in this paper was self-consistently obtained using the SIESTA implementation\cite{Soler2002} of density functional theory (DFT). SIESTA employs norm-conserving pseudo-potentials to account for the core electrons and linear combinations of atomic orbitals to construct the valence states. The generalized gradient approximation (GGA) of the exchange and correlation functional is used with the Perdew-Burke-Ernzerhof parameterization (PBE) a double-zeta polarized (DZP) basis set, a real-space grid defined with an equivalent energy cut-off of 250 Ry. The geometry optimization for each structure is performed to the forces smaller than 20 meV/\AA. 
For the electron band structure calculation, the structure was sampled by a 1 × 1 × 20 Monkhorst–Pack k-point grid and assigned periodic boundary conditions in the (z) direction. For phonon band structre, we first displace each atom by 0.01\AA from their relaxed geometry in the positive and nagative x, y and z directions and calculate force matrix for each geometry. We use Vibra package of Siesta to calclate phonon band structure form the constructed force matrises.
To calculate electron transmission coefficient T(E), the mean-field Hamiltonian of 9-AGNRs with different lengths between CNT electrodes obtained from the converged DFT calculation and combined with Gollum\cite{Sadeghi2015ElectronHeatTransport,Ferrer2014} implementation of the non-equilibrium Green's function method.

\bibliography{references}

\providecommand{\latin}[1]{#1}
\makeatletter
\providecommand{\doi}
  {\begingroup\let\do\@makeother\dospecials
  \catcode`\{=1 \catcode`\}=2 \doi@aux}
\providecommand{\doi@aux}[1]{\endgroup\texttt{#1}}
\makeatother
\providecommand*\mcitethebibliography{\thebibliography}
\csname @ifundefined\endcsname{endmcitethebibliography}
  {\let\endmcitethebibliography\endthebibliography}{}
\begin{mcitethebibliography}{55}
\providecommand*\natexlab[1]{#1}
\providecommand*\mciteSetBstSublistMode[1]{}
\providecommand*\mciteSetBstMaxWidthForm[2]{}
\providecommand*\mciteBstWouldAddEndPuncttrue
  {\def\EndOfBibitem{\unskip.}}
\providecommand*\mciteBstWouldAddEndPunctfalse
  {\let\EndOfBibitem\relax}
\providecommand*\mciteSetBstMidEndSepPunct[3]{}
\providecommand*\mciteSetBstSublistLabelBeginEnd[3]{}
\providecommand*\EndOfBibitem{}
\mciteSetBstSublistMode{f}
\mciteSetBstMaxWidthForm{subitem}{(\alph{mcitesubitemcount})}
\mciteSetBstSublistLabelBeginEnd
  {\mcitemaxwidthsubitemform\space}
  {\relax}
  {\relax}

\bibitem[Ruffieux \latin{et~al.}({2016})Ruffieux, Wang, Yang, Sanchez-Sanchez,
  Liu, Dienel, Talirz, Shinde, Pignedoli, Passerone, Dumslaff, Feng, Muellen,
  and Fasel]{Ruffieux2016surfaceSynthesisGraphene}
Ruffieux,~P.; Wang,~S.; Yang,~B.; Sanchez-Sanchez,~C.; Liu,~J.; Dienel,~T.;
  Talirz,~L.; Shinde,~P.; Pignedoli,~C.~A.; Passerone,~D.; Dumslaff,~T.;
  Feng,~X.; Muellen,~K.; Fasel,~R. On-surface Synthesis of Graphene Nanoribbons
  with Zigzag Edge Topology. \emph{Nature} \textbf{{2016}}, \emph{{531}},
  {489}\relax
\mciteBstWouldAddEndPuncttrue
\mciteSetBstMidEndSepPunct{\mcitedefaultmidpunct}
{\mcitedefaultendpunct}{\mcitedefaultseppunct}\relax
\EndOfBibitem
\bibitem[Cai \latin{et~al.}(2010)Cai, Ruffieux, Jaafar, Bieri, Braun,
  Blankenburg, Muoth, Seitsonen, Saleh, Feng, \latin{et~al.}
  others]{cai2010atomically}
Cai,~J.; Ruffieux,~P.; Jaafar,~R.; Bieri,~M.; Braun,~T.; Blankenburg,~S.;
  Muoth,~M.; Seitsonen,~A.~P.; Saleh,~M.; Feng,~X., \latin{et~al.}  Atomically
  precise bottom-up fabrication of graphene nanoribbons. \emph{Nature}
  \textbf{2010}, \emph{466}, 470--473\relax
\mciteBstWouldAddEndPuncttrue
\mciteSetBstMidEndSepPunct{\mcitedefaultmidpunct}
{\mcitedefaultendpunct}{\mcitedefaultseppunct}\relax
\EndOfBibitem
\bibitem[Groning \latin{et~al.}({2018})Groning, Wang, Yao, Pignedoli, Barin,
  Daniels, Cupo, Meunier, Feng, Narita, Muellen, Ruffieux, and
  Fasel]{Groning2018EngineeringRobustTopological}
Groning,~O.; Wang,~S.; Yao,~X.; Pignedoli,~C.~A.; Barin,~G.~B.; Daniels,~C.;
  Cupo,~A.; Meunier,~V.; Feng,~X.; Narita,~A.; Muellen,~K.; Ruffieux,~P.;
  Fasel,~R. Engineering of Robust Topological Quantum Phases in Graphene
  Nanoribbons. \emph{Nature} \textbf{{2018}}, \emph{{560}}, {209}\relax
\mciteBstWouldAddEndPuncttrue
\mciteSetBstMidEndSepPunct{\mcitedefaultmidpunct}
{\mcitedefaultendpunct}{\mcitedefaultseppunct}\relax
\EndOfBibitem
\bibitem[Rizzo \latin{et~al.}(2018)Rizzo, Veber, Cao, Bronner, Chen, Zhao,
  Rodriguez, Louie, Crommie, and Fischer]{Rizzo2018TopologicalBandEngineering}
Rizzo,~D.~J.; Veber,~G.; Cao,~T.; Bronner,~C.; Chen,~T.; Zhao,~F.;
  Rodriguez,~H.; Louie,~S.~G.; Crommie,~M.~F.; Fischer,~F.~R. Topological Band
  Engineering of Graphene Nanoribbons. \emph{Nature} \textbf{2018}, \emph{560},
  204--208\relax
\mciteBstWouldAddEndPuncttrue
\mciteSetBstMidEndSepPunct{\mcitedefaultmidpunct}
{\mcitedefaultendpunct}{\mcitedefaultseppunct}\relax
\EndOfBibitem
\bibitem[Llinas \latin{et~al.}(2017)Llinas, Fairbrother, Borin~Barin, Shi, Lee,
  Wu, Yong~Choi, Braganza, Lear, Kau, \latin{et~al.} others]{llinas2017short}
Llinas,~J.~P.; Fairbrother,~A.; Borin~Barin,~G.; Shi,~W.; Lee,~K.; Wu,~S.;
  Yong~Choi,~B.; Braganza,~R.; Lear,~J.; Kau,~N., \latin{et~al.}  Short-channel
  field-effect transistors with 9-atom and 13-atom wide graphene nanoribbons.
  \emph{Nature communications} \textbf{2017}, \emph{8}, 1--6\relax
\mciteBstWouldAddEndPuncttrue
\mciteSetBstMidEndSepPunct{\mcitedefaultmidpunct}
{\mcitedefaultendpunct}{\mcitedefaultseppunct}\relax
\EndOfBibitem
\bibitem[Bennett \latin{et~al.}(2013)Bennett, Pedramrazi, Madani, Chen,
  de~Oteyza, Chen, Fischer, Crommie, and Bokor]{bennett2013bottom}
Bennett,~P.~B.; Pedramrazi,~Z.; Madani,~A.; Chen,~Y.-C.; de~Oteyza,~D.~G.;
  Chen,~C.; Fischer,~F.~R.; Crommie,~M.~F.; Bokor,~J. Bottom-up graphene
  nanoribbon field-effect transistors. \emph{Applied Physics Letters}
  \textbf{2013}, \emph{103}, 253114\relax
\mciteBstWouldAddEndPuncttrue
\mciteSetBstMidEndSepPunct{\mcitedefaultmidpunct}
{\mcitedefaultendpunct}{\mcitedefaultseppunct}\relax
\EndOfBibitem
\bibitem[Richter \latin{et~al.}(2020)Richter, Chen, Tries, Prechtl, Narita,
  M{\"u}llen, Asadi, Bonn, and Kl{\"a}ui]{richter2020charge}
Richter,~N.; Chen,~Z.; Tries,~A.; Prechtl,~T.; Narita,~A.; M{\"u}llen,~K.;
  Asadi,~K.; Bonn,~M.; Kl{\"a}ui,~M. Charge transport mechanism in networks of
  armchair graphene nanoribbons. \emph{Scientific reports} \textbf{2020},
  \emph{10}, 1--8\relax
\mciteBstWouldAddEndPuncttrue
\mciteSetBstMidEndSepPunct{\mcitedefaultmidpunct}
{\mcitedefaultendpunct}{\mcitedefaultseppunct}\relax
\EndOfBibitem
\bibitem[Mutlu \latin{et~al.}(2021)Mutlu, Lin, Barin, Zhang, Pitner, Wang,
  Darawish, Di~Giovannantonio, Wang, Cai, \latin{et~al.}
  others]{mutlu2021short}
Mutlu,~Z.; Lin,~Y.; Barin,~G.; Zhang,~Z.; Pitner,~G.; Wang,~S.; Darawish,~R.;
  Di~Giovannantonio,~M.; Wang,~H.; Cai,~J., \latin{et~al.}  Short-Channel
  Double-Gate FETs with Atomically Precise Graphene Nanoribbons. 2021 IEEE
  International Electron Devices Meeting (IEDM). 2021; pp 37--4\relax
\mciteBstWouldAddEndPuncttrue
\mciteSetBstMidEndSepPunct{\mcitedefaultmidpunct}
{\mcitedefaultendpunct}{\mcitedefaultseppunct}\relax
\EndOfBibitem
\bibitem[Wang \latin{et~al.}(2021)Wang, Chen, Elibol, He, Wang, Chen, Jiang,
  Li, Wu, Cong, \latin{et~al.} others]{wang2021towards}
Wang,~H.~S.; Chen,~L.; Elibol,~K.; He,~L.; Wang,~H.; Chen,~C.; Jiang,~C.;
  Li,~C.; Wu,~T.; Cong,~C.~X., \latin{et~al.}  Towards chirality control of
  graphene nanoribbons embedded in hexagonal boron nitride. \emph{Nature
  Materials} \textbf{2021}, \emph{20}, 202--207\relax
\mciteBstWouldAddEndPuncttrue
\mciteSetBstMidEndSepPunct{\mcitedefaultmidpunct}
{\mcitedefaultendpunct}{\mcitedefaultseppunct}\relax
\EndOfBibitem
\bibitem[Senkovskiy \latin{et~al.}(2021)Senkovskiy, Nenashev, Alavi, Falke,
  Hell, Bampoulis, Rybkovskiy, Usachov, Fedorov, Chernov, \latin{et~al.}
  others]{senkovskiy2021tunneling}
Senkovskiy,~B.~V.; Nenashev,~A.~V.; Alavi,~S.~K.; Falke,~Y.; Hell,~M.;
  Bampoulis,~P.; Rybkovskiy,~D.~V.; Usachov,~D.~Y.; Fedorov,~A.~V.;
  Chernov,~A.~I., \latin{et~al.}  Tunneling current modulation in atomically
  precise graphene nanoribbon heterojunctions. \emph{Nature communications}
  \textbf{2021}, \emph{12}, 1--11\relax
\mciteBstWouldAddEndPuncttrue
\mciteSetBstMidEndSepPunct{\mcitedefaultmidpunct}
{\mcitedefaultendpunct}{\mcitedefaultseppunct}\relax
\EndOfBibitem
\bibitem[Sakaguchi \latin{et~al.}(2017)Sakaguchi, Song, Kojima, and
  Nakae]{sakaguchi2017homochiral}
Sakaguchi,~H.; Song,~S.; Kojima,~T.; Nakae,~T. Homochiral polymerization-driven
  selective growth of graphene nanoribbons. \emph{Nature Chemistry}
  \textbf{2017}, \emph{9}, 57--63\relax
\mciteBstWouldAddEndPuncttrue
\mciteSetBstMidEndSepPunct{\mcitedefaultmidpunct}
{\mcitedefaultendpunct}{\mcitedefaultseppunct}\relax
\EndOfBibitem
\bibitem[Mutlu \latin{et~al.}(2021)Mutlu, Llinas, Jacobse, Piskun, Blackwell,
  Crommie, Fischer, and Bokor]{mutlu2021transfer}
Mutlu,~Z.; Llinas,~J.~P.; Jacobse,~P.~H.; Piskun,~I.; Blackwell,~R.;
  Crommie,~M.~F.; Fischer,~F.~R.; Bokor,~J. Transfer-Free Synthesis of
  Atomically Precise Graphene Nanoribbons on Insulating Substrates. \emph{ACS
  nano} \textbf{2021}, \emph{15}, 2635--2642\relax
\mciteBstWouldAddEndPuncttrue
\mciteSetBstMidEndSepPunct{\mcitedefaultmidpunct}
{\mcitedefaultendpunct}{\mcitedefaultseppunct}\relax
\EndOfBibitem
\bibitem[Jacobberger \latin{et~al.}(2015)Jacobberger, Kiraly, Fortin-Deschenes,
  Levesque, McElhinny, Brady, Rojas~Delgado, Singha~Roy, Mannix, Lagally,
  \latin{et~al.} others]{jacobberger2015direct}
Jacobberger,~R.~M.; Kiraly,~B.; Fortin-Deschenes,~M.; Levesque,~P.~L.;
  McElhinny,~K.~M.; Brady,~G.~J.; Rojas~Delgado,~R.; Singha~Roy,~S.;
  Mannix,~A.; Lagally,~M.~G., \latin{et~al.}  Direct oriented growth of
  armchair graphene nanoribbons on germanium. \emph{Nature communications}
  \textbf{2015}, \emph{6}, 1--8\relax
\mciteBstWouldAddEndPuncttrue
\mciteSetBstMidEndSepPunct{\mcitedefaultmidpunct}
{\mcitedefaultendpunct}{\mcitedefaultseppunct}\relax
\EndOfBibitem
\bibitem[Way \latin{et~al.}(2018)Way, Jacobberger, and Arnold]{way2018seed}
Way,~A.~J.; Jacobberger,~R.~M.; Arnold,~M.~S. Seed-initiated anisotropic growth
  of unidirectional armchair graphene nanoribbon arrays on germanium.
  \emph{Nano letters} \textbf{2018}, \emph{18}, 898--906\relax
\mciteBstWouldAddEndPuncttrue
\mciteSetBstMidEndSepPunct{\mcitedefaultmidpunct}
{\mcitedefaultendpunct}{\mcitedefaultseppunct}\relax
\EndOfBibitem
\bibitem[Shekhirev \latin{et~al.}(2017)Shekhirev, Vo, Mehdi~Pour, Lipatov,
  Munukutla, Lyding, and Sinitskii]{shekhirev2017interfacial}
Shekhirev,~M.; Vo,~T.~H.; Mehdi~Pour,~M.; Lipatov,~A.; Munukutla,~S.;
  Lyding,~J.~W.; Sinitskii,~A. Interfacial self-assembly of atomically precise
  graphene nanoribbons into uniform thin films for electronics applications.
  \emph{ACS Applied Materials \& Interfaces} \textbf{2017}, \emph{9},
  693--700\relax
\mciteBstWouldAddEndPuncttrue
\mciteSetBstMidEndSepPunct{\mcitedefaultmidpunct}
{\mcitedefaultendpunct}{\mcitedefaultseppunct}\relax
\EndOfBibitem
\bibitem[Braun \latin{et~al.}(2021)Braun, Overbeck, El~Abbassi, K{\"a}ser,
  Furrer, Olziersky, Flasby, Barin, Sun, Darawish, \latin{et~al.}
  others]{braun2021optimized}
Braun,~O.; Overbeck,~J.; El~Abbassi,~M.; K{\"a}ser,~S.; Furrer,~R.;
  Olziersky,~A.; Flasby,~A.; Barin,~G.~B.; Sun,~Q.; Darawish,~R.,
  \latin{et~al.}  Optimized graphene electrodes for contacting graphene
  nanoribbons. \emph{Carbon} \textbf{2021}, \emph{184}, 331--339\relax
\mciteBstWouldAddEndPuncttrue
\mciteSetBstMidEndSepPunct{\mcitedefaultmidpunct}
{\mcitedefaultendpunct}{\mcitedefaultseppunct}\relax
\EndOfBibitem
\bibitem[Chen \latin{et~al.}(2016)Chen, Zhang, Palma, Lodi~Rizzini, Liu, Abbas,
  Richter, Martini, Wang, Cavani, \latin{et~al.} others]{chen2016synthesis}
Chen,~Z.; Zhang,~W.; Palma,~C.-A.; Lodi~Rizzini,~A.; Liu,~B.; Abbas,~A.;
  Richter,~N.; Martini,~L.; Wang,~X.-Y.; Cavani,~N., \latin{et~al.}  Synthesis
  of graphene nanoribbons by ambient-pressure chemical vapor deposition and
  device integration. \emph{Journal of the American Chemical Society}
  \textbf{2016}, \emph{138}, 15488--15496\relax
\mciteBstWouldAddEndPuncttrue
\mciteSetBstMidEndSepPunct{\mcitedefaultmidpunct}
{\mcitedefaultendpunct}{\mcitedefaultseppunct}\relax
\EndOfBibitem
\bibitem[Candini \latin{et~al.}(2017)Candini, Martini, Chen, Mishra,
  Convertino, Coletti, Narita, Feng, M\"ullen, and Affronte]{candini2017high}
Candini,~A.; Martini,~L.; Chen,~Z.; Mishra,~N.; Convertino,~D.; Coletti,~C.;
  Narita,~A.; Feng,~X.; M\"ullen,~K.; Affronte,~M. High photoresponsivity in
  graphene nanoribbon field-effect transistor devices contacted with graphene
  electrodes. \emph{The Journal of Physical Chemistry C} \textbf{2017},
  \emph{121}, 10620--10625\relax
\mciteBstWouldAddEndPuncttrue
\mciteSetBstMidEndSepPunct{\mcitedefaultmidpunct}
{\mcitedefaultendpunct}{\mcitedefaultseppunct}\relax
\EndOfBibitem
\bibitem[El~Abbassi \latin{et~al.}(2020)El~Abbassi, Perrin, Barin, Sangtarash,
  Overbeck, Braun, Lambert, Sun, Prechtl, Narita, \latin{et~al.}
  others]{el2020controlled}
El~Abbassi,~M.; Perrin,~M.~L.; Barin,~G.~B.; Sangtarash,~S.; Overbeck,~J.;
  Braun,~O.; Lambert,~C.~J.; Sun,~Q.; Prechtl,~T.; Narita,~A., \latin{et~al.}
  Controlled quantum dot formation in atomically engineered graphene nanoribbon
  field-effect transistors. \emph{ACS nano} \textbf{2020}, \emph{14},
  5754--5762\relax
\mciteBstWouldAddEndPuncttrue
\mciteSetBstMidEndSepPunct{\mcitedefaultmidpunct}
{\mcitedefaultendpunct}{\mcitedefaultseppunct}\relax
\EndOfBibitem
\bibitem[Sun \latin{et~al.}(2020)Sun, Gr{\"o}ning, Overbeck, Braun, Perrin,
  Borin~Barin, El~Abbassi, Eimre, Ditler, Daniels, \latin{et~al.}
  others]{sun2020massive}
Sun,~Q.; Gr{\"o}ning,~O.; Overbeck,~J.; Braun,~O.; Perrin,~M.~L.;
  Borin~Barin,~G.; El~Abbassi,~M.; Eimre,~K.; Ditler,~E.; Daniels,~C.,
  \latin{et~al.}  Massive Dirac fermion behavior in a low bandgap graphene
  nanoribbon near a topological phase boundary. \emph{Advanced materials}
  \textbf{2020}, \emph{32}, 1906054\relax
\mciteBstWouldAddEndPuncttrue
\mciteSetBstMidEndSepPunct{\mcitedefaultmidpunct}
{\mcitedefaultendpunct}{\mcitedefaultseppunct}\relax
\EndOfBibitem
\bibitem[Martini \latin{et~al.}(2019)Martini, Chen, Mishra, Barin, Fantuzzi,
  Ruffieux, Fasel, Feng, Narita, Coletti, \latin{et~al.}
  others]{martini2019structure}
Martini,~L.; Chen,~Z.; Mishra,~N.; Barin,~G.~B.; Fantuzzi,~P.; Ruffieux,~P.;
  Fasel,~R.; Feng,~X.; Narita,~A.; Coletti,~C., \latin{et~al.}
  Structure-dependent electrical properties of graphene nanoribbon devices with
  graphene electrodes. \emph{Carbon} \textbf{2019}, \emph{146}, 36--43\relax
\mciteBstWouldAddEndPuncttrue
\mciteSetBstMidEndSepPunct{\mcitedefaultmidpunct}
{\mcitedefaultendpunct}{\mcitedefaultseppunct}\relax
\EndOfBibitem
\bibitem[Fantuzzi \latin{et~al.}(2016)Fantuzzi, Martini, Candini, Corradini,
  Del~Pennino, Hu, Feng, M{\"u}llen, Narita, and
  Affronte]{fantuzzi2016fabrication}
Fantuzzi,~P.; Martini,~L.; Candini,~A.; Corradini,~V.; Del~Pennino,~U.; Hu,~Y.;
  Feng,~X.; M{\"u}llen,~K.; Narita,~A.; Affronte,~M. Fabrication of three
  terminal devices by ElectroSpray deposition of graphene nanoribbons.
  \emph{Carbon} \textbf{2016}, \emph{104}, 112--118\relax
\mciteBstWouldAddEndPuncttrue
\mciteSetBstMidEndSepPunct{\mcitedefaultmidpunct}
{\mcitedefaultendpunct}{\mcitedefaultseppunct}\relax
\EndOfBibitem
\bibitem[Qiu \latin{et~al.}(2017)Qiu, Zhang, Xiao, Yang, Zhong, and
  Peng]{qiu2017scaling}
Qiu,~C.; Zhang,~Z.; Xiao,~M.; Yang,~Y.; Zhong,~D.; Peng,~L.-M. Scaling carbon
  nanotube complementary transistors to 5-nm gate lengths. \emph{Science}
  \textbf{2017}, \emph{355}, 271--276\relax
\mciteBstWouldAddEndPuncttrue
\mciteSetBstMidEndSepPunct{\mcitedefaultmidpunct}
{\mcitedefaultendpunct}{\mcitedefaultseppunct}\relax
\EndOfBibitem
\bibitem[Chen \latin{et~al.}(2021)Chen, Shu, Zhang, Zeng, Liang, Zheng, and
  Duan]{chen2021sub}
Chen,~Y.; Shu,~Z.; Zhang,~S.; Zeng,~P.; Liang,~H.; Zheng,~M.; Duan,~H.
  Sub-10-nm fabrication: Methods and applications. \emph{International Journal
  of Extreme Manufacturing} \textbf{2021}, \relax
\mciteBstWouldAddEndPunctfalse
\mciteSetBstMidEndSepPunct{\mcitedefaultmidpunct}
{}{\mcitedefaultseppunct}\relax
\EndOfBibitem
\bibitem[Lin \latin{et~al.}(2022)Lin, Mutlu, Barin, Hong, Llinas, Narita,
  Singh, M{\"u}llen, Ruffieux, Fasel, \latin{et~al.} others]{lin2022scaling}
Lin,~Y.; Mutlu,~Z.; Barin,~G.~B.; Hong,~J.; Llinas,~J.~P.; Narita,~A.;
  Singh,~H.; M{\"u}llen,~K.; Ruffieux,~P.; Fasel,~R., \latin{et~al.}  Scaling
  and Statistics of Bottom-Up Synthesized Armchair Graphene Nanoribbon
  Transistors. \emph{arXiv preprint arXiv:2201.09341} \textbf{2022}, \relax
\mciteBstWouldAddEndPunctfalse
\mciteSetBstMidEndSepPunct{\mcitedefaultmidpunct}
{}{\mcitedefaultseppunct}\relax
\EndOfBibitem
\bibitem[{El Abbassi} \latin{et~al.}(2020){El Abbassi}, Perrin, Barin,
  Sangtarash, Overbeck, Braun, Lambert, Sun, Prechtl, Narita, M{\"{u}}llen,
  Ruffieux, Sadeghi, Fasel, and Calame]{ElAbbassi2020ControlledQuantumDot}
{El Abbassi},~M.; Perrin,~M.~L.; Barin,~G.~B.; Sangtarash,~S.; Overbeck,~J.;
  Braun,~O.; Lambert,~C.~J.; Sun,~Q.; Prechtl,~T.; Narita,~A.;
  M{\"{u}}llen,~K.; Ruffieux,~P.; Sadeghi,~H.; Fasel,~R.; Calame,~M. Controlled
  Quantum Dot Formation in Atomically Engineered Graphene Nanoribbon
  Field-effect Transistors. \emph{{{ACS Nano}}} \textbf{2020}, \emph{14},
  5754--5762\relax
\mciteBstWouldAddEndPuncttrue
\mciteSetBstMidEndSepPunct{\mcitedefaultmidpunct}
{\mcitedefaultendpunct}{\mcitedefaultseppunct}\relax
\EndOfBibitem
\bibitem[Mishra \latin{et~al.}(2021)Mishra, Catarina, Wu, Ortiz, Jacob, Eimre,
  Ma, Pignedoli, Feng, Ruffieux, \latin{et~al.} others]{mishra2021observation}
Mishra,~S.; Catarina,~G.; Wu,~F.; Ortiz,~R.; Jacob,~D.; Eimre,~K.; Ma,~J.;
  Pignedoli,~C.~A.; Feng,~X.; Ruffieux,~P., \latin{et~al.}  Observation of
  fractional edge excitations in nanographene spin chains. \emph{Nature}
  \textbf{2021}, \emph{598}, 287--292\relax
\mciteBstWouldAddEndPuncttrue
\mciteSetBstMidEndSepPunct{\mcitedefaultmidpunct}
{\mcitedefaultendpunct}{\mcitedefaultseppunct}\relax
\EndOfBibitem
\bibitem[Gr{\"{o}}ning \latin{et~al.}(2018)Gr{\"{o}}ning, Wang, Yao, Pignedoli,
  {Borin Barin}, Daniels, Cupo, Meunier, Feng, Narita, M{\"{u}}llen, Ruffieux,
  and Fasel]{Groning2018}
Gr{\"{o}}ning,~O.; Wang,~S.; Yao,~X.; Pignedoli,~C.~A.; {Borin Barin},~G.;
  Daniels,~C.; Cupo,~A.; Meunier,~V.; Feng,~X.; Narita,~A.; M{\"{u}}llen,~K.;
  Ruffieux,~P.; Fasel,~R. {Engineering of robust topological quantum phases in
  graphene nanoribbons}. \emph{Nature} \textbf{2018}, \emph{560},
  209--213\relax
\mciteBstWouldAddEndPuncttrue
\mciteSetBstMidEndSepPunct{\mcitedefaultmidpunct}
{\mcitedefaultendpunct}{\mcitedefaultseppunct}\relax
\EndOfBibitem
\bibitem[Mintmire and White(1998)Mintmire, and White]{mintmire1998universal}
Mintmire,~J.; White,~C. Universal density of states for carbon nanotubes.
  \emph{Physical Review Letters} \textbf{1998}, \emph{81}, 2506\relax
\mciteBstWouldAddEndPuncttrue
\mciteSetBstMidEndSepPunct{\mcitedefaultmidpunct}
{\mcitedefaultendpunct}{\mcitedefaultseppunct}\relax
\EndOfBibitem
\bibitem[Odom \latin{et~al.}(2000)Odom, Huang, Kim, and
  Lieber]{odom2000structure}
Odom,~T.~W.; Huang,~J.-L.; Kim,~P.; Lieber,~C.~M. Structure and electronic
  properties of carbon nanotubes. 2000\relax
\mciteBstWouldAddEndPuncttrue
\mciteSetBstMidEndSepPunct{\mcitedefaultmidpunct}
{\mcitedefaultendpunct}{\mcitedefaultseppunct}\relax
\EndOfBibitem
\bibitem[Richter \latin{et~al.}(2020)Richter, Chen, Tries, Prechtl, Narita,
  M{\"{u}}llen, Asadi, Bonn, and
  Kl{\"{a}}ui]{Richter2020ChargeTransportMechanism}
Richter,~N.; Chen,~Z.; Tries,~A.; Prechtl,~T.; Narita,~A.; M{\"{u}}llen,~K.;
  Asadi,~K.; Bonn,~M.; Kl{\"{a}}ui,~M. Charge Transport Mechanism in Networks
  of Armchair Graphene Nanoribbons. \emph{Scientific Reports} \textbf{2020},
  \emph{10}, 1--8\relax
\mciteBstWouldAddEndPuncttrue
\mciteSetBstMidEndSepPunct{\mcitedefaultmidpunct}
{\mcitedefaultendpunct}{\mcitedefaultseppunct}\relax
\EndOfBibitem
\bibitem[Thomas \latin{et~al.}(2021)Thomas, Nilsson, Ciaccia, J{\"u}nger,
  Rossi, Zannier, Sorba, Baumgartner, and
  Sch{\"o}nenberger]{thomas2021spectroscopy}
Thomas,~F.~S.; Nilsson,~M.; Ciaccia,~C.; J{\"u}nger,~C.; Rossi,~F.;
  Zannier,~V.; Sorba,~L.; Baumgartner,~A.; Sch{\"o}nenberger,~C. Spectroscopy
  of the local density of states in nanowires using integrated quantum dots.
  \emph{Physical Review B} \textbf{2021}, \emph{104}, 115415\relax
\mciteBstWouldAddEndPuncttrue
\mciteSetBstMidEndSepPunct{\mcitedefaultmidpunct}
{\mcitedefaultendpunct}{\mcitedefaultseppunct}\relax
\EndOfBibitem
\bibitem[G{\"u}ttinger \latin{et~al.}(2009)G{\"u}ttinger, Stampfer, Libisch,
  Frey, Burgd{\"o}rfer, Ihn, and Ensslin]{guttinger2009electron}
G{\"u}ttinger,~J.; Stampfer,~C.; Libisch,~F.; Frey,~T.; Burgd{\"o}rfer,~J.;
  Ihn,~T.; Ensslin,~K. Electron-hole crossover in graphene quantum dots.
  \emph{Physical review letters} \textbf{2009}, \emph{103}, 046810\relax
\mciteBstWouldAddEndPuncttrue
\mciteSetBstMidEndSepPunct{\mcitedefaultmidpunct}
{\mcitedefaultendpunct}{\mcitedefaultseppunct}\relax
\EndOfBibitem
\bibitem[Park \latin{et~al.}(2000)Park, Park, Lim, Anderson, Alivisatos, and
  McEuen]{park2000nanomechanical}
Park,~H.; Park,~J.; Lim,~A.~K.; Anderson,~E.~H.; Alivisatos,~A.~P.;
  McEuen,~P.~L. Nanomechanical oscillations in a single-C60 transistor.
  \emph{Nature} \textbf{2000}, \emph{407}, 57--60\relax
\mciteBstWouldAddEndPuncttrue
\mciteSetBstMidEndSepPunct{\mcitedefaultmidpunct}
{\mcitedefaultendpunct}{\mcitedefaultseppunct}\relax
\EndOfBibitem
\bibitem[Lau \latin{et~al.}(2016)Lau, Sadeghi, Rogers, Sangtarash, Dallas,
  Porfyrakis, Warner, Lambert, Briggs, and Mol]{lau2016redox}
Lau,~C.~S.; Sadeghi,~H.; Rogers,~G.; Sangtarash,~S.; Dallas,~P.;
  Porfyrakis,~K.; Warner,~J.; Lambert,~C.~J.; Briggs,~G. A.~D.; Mol,~J.~A.
  Redox-dependent Franck--Condon blockade and Avalanche transport in a
  graphene--fullerene single-molecule transistor. \emph{Nano letters}
  \textbf{2016}, \emph{16}, 170--176\relax
\mciteBstWouldAddEndPuncttrue
\mciteSetBstMidEndSepPunct{\mcitedefaultmidpunct}
{\mcitedefaultendpunct}{\mcitedefaultseppunct}\relax
\EndOfBibitem
\bibitem[Burzur{\'\i} \latin{et~al.}(2014)Burzur{\'\i}, Yamamoto, Warnock,
  Zhong, Park, Cornia, and van~der Zant]{burzuri2014franck}
Burzur{\'\i},~E.; Yamamoto,~Y.; Warnock,~M.; Zhong,~X.; Park,~K.; Cornia,~A.;
  van~der Zant,~H.~S. Franck--Condon blockade in a single-molecule transistor.
  \emph{Nano letters} \textbf{2014}, \emph{14}, 3191--3196\relax
\mciteBstWouldAddEndPuncttrue
\mciteSetBstMidEndSepPunct{\mcitedefaultmidpunct}
{\mcitedefaultendpunct}{\mcitedefaultseppunct}\relax
\EndOfBibitem
\bibitem[Pasupathy \latin{et~al.}(2005)Pasupathy, Park, Chang, Soldatov,
  Lebedkin, Bialczak, Grose, Donev, Sethna, Ralph, \latin{et~al.}
  others]{pasupathy2005vibration}
Pasupathy,~A.; Park,~J.; Chang,~C.; Soldatov,~A.; Lebedkin,~S.; Bialczak,~R.;
  Grose,~J.; Donev,~L.; Sethna,~J.; Ralph,~D., \latin{et~al.}
  Vibration-assisted electron tunneling in C140 transistors. \emph{Nano
  Letters} \textbf{2005}, \emph{5}, 203--207\relax
\mciteBstWouldAddEndPuncttrue
\mciteSetBstMidEndSepPunct{\mcitedefaultmidpunct}
{\mcitedefaultendpunct}{\mcitedefaultseppunct}\relax
\EndOfBibitem
\bibitem[Sapmaz \latin{et~al.}(2006)Sapmaz, Jarillo-Herrero, Blanter, Dekker,
  and Van Der~Zant]{sapmaz2006tunneling}
Sapmaz,~S.; Jarillo-Herrero,~P.; Blanter,~Y.~M.; Dekker,~C.; Van Der~Zant,~H.
  Tunneling in suspended carbon nanotubes assisted by longitudinal phonons.
  \emph{Physical review letters} \textbf{2006}, \emph{96}, 026801\relax
\mciteBstWouldAddEndPuncttrue
\mciteSetBstMidEndSepPunct{\mcitedefaultmidpunct}
{\mcitedefaultendpunct}{\mcitedefaultseppunct}\relax
\EndOfBibitem
\bibitem[Leturcq \latin{et~al.}(2009)Leturcq, Stampfer, Inderbitzin, Durrer,
  Hierold, Mariani, Schultz, Von~Oppen, and Ensslin]{leturcq2009franck}
Leturcq,~R.; Stampfer,~C.; Inderbitzin,~K.; Durrer,~L.; Hierold,~C.;
  Mariani,~E.; Schultz,~M.~G.; Von~Oppen,~F.; Ensslin,~K. Franck--Condon
  blockade in suspended carbon nanotube quantum dots. \emph{Nature Physics}
  \textbf{2009}, \emph{5}, 327--331\relax
\mciteBstWouldAddEndPuncttrue
\mciteSetBstMidEndSepPunct{\mcitedefaultmidpunct}
{\mcitedefaultendpunct}{\mcitedefaultseppunct}\relax
\EndOfBibitem
\bibitem[Weig \latin{et~al.}(2004)Weig, Blick, Brandes, Kirschbaum,
  Wegscheider, Bichler, and Kotthaus]{weig2004single}
Weig,~E.~M.; Blick,~R.~H.; Brandes,~T.; Kirschbaum,~J.; Wegscheider,~W.;
  Bichler,~M.; Kotthaus,~J.~P. Single-electron-phonon interaction in a
  suspended quantum dot phonon cavity. \emph{Physical review letters}
  \textbf{2004}, \emph{92}, 046804\relax
\mciteBstWouldAddEndPuncttrue
\mciteSetBstMidEndSepPunct{\mcitedefaultmidpunct}
{\mcitedefaultendpunct}{\mcitedefaultseppunct}\relax
\EndOfBibitem
\bibitem[Nowack and Wegewijs(2005)Nowack, and Wegewijs]{nowack2005vibration}
Nowack,~K.~C.; Wegewijs,~M.~R. Vibration-assisted tunneling through competing
  molecular states. \emph{arXiv preprint cond-mat/0506552} \textbf{2005},
  \relax
\mciteBstWouldAddEndPunctfalse
\mciteSetBstMidEndSepPunct{\mcitedefaultmidpunct}
{}{\mcitedefaultseppunct}\relax
\EndOfBibitem
\bibitem[McCarthy \latin{et~al.}(2003)McCarthy, Prokof’ev, and
  Tuominen]{mccarthy2003incoherent}
McCarthy,~K.~D.; Prokof’ev,~N.; Tuominen,~M.~T. Incoherent dynamics of
  vibrating single-molecule transistors. \emph{Physical Review B}
  \textbf{2003}, \emph{67}, 245415\relax
\mciteBstWouldAddEndPuncttrue
\mciteSetBstMidEndSepPunct{\mcitedefaultmidpunct}
{\mcitedefaultendpunct}{\mcitedefaultseppunct}\relax
\EndOfBibitem
\bibitem[Koch and Von~Oppen(2005)Koch, and Von~Oppen]{koch2005franck}
Koch,~J.; Von~Oppen,~F. Franck-Condon blockade and giant Fano factors in
  transport through single molecules. \emph{Physical review letters}
  \textbf{2005}, \emph{94}, 206804\relax
\mciteBstWouldAddEndPuncttrue
\mciteSetBstMidEndSepPunct{\mcitedefaultmidpunct}
{\mcitedefaultendpunct}{\mcitedefaultseppunct}\relax
\EndOfBibitem
\bibitem[Sadeghi(2018)]{Sadeghi2018TheoryElectronPhonon}
Sadeghi,~H. Theory of Electron, Phonon and Spin Transport in Nanoscale Quantum
  Devices. \emph{Nanotechnology} \textbf{2018}, \emph{29}, 373001\relax
\mciteBstWouldAddEndPuncttrue
\mciteSetBstMidEndSepPunct{\mcitedefaultmidpunct}
{\mcitedefaultendpunct}{\mcitedefaultseppunct}\relax
\EndOfBibitem
\bibitem[Son \latin{et~al.}(2006)Son, Cohen, and
  Louie]{Son2006EnergyGapsGraphene}
Son,~Y.-W.; Cohen,~M.~L.; Louie,~S.~G. Energy Gaps in Graphene Nanoribbons.
  \emph{Physical Review Letters} \textbf{2006}, \emph{97}, 216803\relax
\mciteBstWouldAddEndPuncttrue
\mciteSetBstMidEndSepPunct{\mcitedefaultmidpunct}
{\mcitedefaultendpunct}{\mcitedefaultseppunct}\relax
\EndOfBibitem
\bibitem[Biercuk \latin{et~al.}(2004)Biercuk, Mason, Chow, and
  Marcus]{biercuk2004locally}
Biercuk,~M.; Mason,~N.; Chow,~J.; Marcus,~C. Locally addressable tunnel
  barriers within a carbon nanotube. \emph{Nano letters} \textbf{2004},
  \emph{4}, 2499--2502\relax
\mciteBstWouldAddEndPuncttrue
\mciteSetBstMidEndSepPunct{\mcitedefaultmidpunct}
{\mcitedefaultendpunct}{\mcitedefaultseppunct}\relax
\EndOfBibitem
\bibitem[Overbeck \latin{et~al.}(2019)Overbeck, {Borin Barin}, Daniels, Perrin,
  Liang, Braun, Darawish, Burkhardt, Dumslaff, Wang, Narita, M{\"{u}}llen,
  Meunier, Fasel, Calame, and
  Ruffieux]{Overbeck2019OptimizedSubstratesMeasurement}
Overbeck,~J.; {Borin Barin},~G.; Daniels,~C.; Perrin,~M.~L.; Liang,~L.;
  Braun,~O.; Darawish,~R.; Burkhardt,~B.; Dumslaff,~T.; Wang,~X.~Y.;
  Narita,~A.; M{\"{u}}llen,~K.; Meunier,~V.; Fasel,~R.; Calame,~M.;
  Ruffieux,~P. Optimized Substrates and Measurement Approaches for Raman
  Spectroscopy of Graphene Nanoribbons. \emph{Physica Status Solidi (b) Basic
  Research} \textbf{2019}, \emph{256}, 1900343\relax
\mciteBstWouldAddEndPuncttrue
\mciteSetBstMidEndSepPunct{\mcitedefaultmidpunct}
{\mcitedefaultendpunct}{\mcitedefaultseppunct}\relax
\EndOfBibitem
\bibitem[Di~Giovannantonio \latin{et~al.}(2018)Di~Giovannantonio, Deniz, Urgel,
  Widmer, Dienel, Stolz, S{\'a}nchez-S{\'a}nchez, Muntwiler, Dumslaff, Berger,
  Narita, Feng, M{\"u}llen, Ruffieux, and
  Fasel]{DiGiovannantonio2018surfaceGrowthDynamics}
Di~Giovannantonio,~M.; Deniz,~O.; Urgel,~J.~I.; Widmer,~R.; Dienel,~T.;
  Stolz,~S.; S{\'a}nchez-S{\'a}nchez,~C.; Muntwiler,~M.; Dumslaff,~T.;
  Berger,~R.; Narita,~A.; Feng,~X.; M{\"u}llen,~K.; Ruffieux,~P.; Fasel,~R.
  On-surface Growth Dynamics of Graphene Nanoribbons: The Role of Halogen
  Functionalization. \emph{{ACS Nano}} \textbf{2018}, \emph{12}, 74--81\relax
\mciteBstWouldAddEndPuncttrue
\mciteSetBstMidEndSepPunct{\mcitedefaultmidpunct}
{\mcitedefaultendpunct}{\mcitedefaultseppunct}\relax
\EndOfBibitem
\bibitem[Senkovskiy \latin{et~al.}(2017)Senkovskiy, Pfeiffer, Alavi, Bliesener,
  Zhu, Michel, Fedorov, German, Hertel, Haberer, Petaccia, Fischer, Meerholz,
  van Loosdrecht, Lindfors, and Grüneis]{senkovskiy2017making}
Senkovskiy,~B.; Pfeiffer,~M.; Alavi,~S.; Bliesener,~A.; Zhu,~J.; Michel,~S.;
  Fedorov,~A.; German,~R.; Hertel,~D.; Haberer,~D.; Petaccia,~L.; Fischer,~F.;
  Meerholz,~K.; van Loosdrecht,~P.; Lindfors,~K.; Grüneis,~A. Making graphene
  nanoribbons photoluminescent. \emph{Nano letters} \textbf{2017}, \emph{17},
  4029--4037\relax
\mciteBstWouldAddEndPuncttrue
\mciteSetBstMidEndSepPunct{\mcitedefaultmidpunct}
{\mcitedefaultendpunct}{\mcitedefaultseppunct}\relax
\EndOfBibitem
\bibitem[Overbeck \latin{et~al.}(2019)Overbeck, Barin, Daniels, Perrin, Braun,
  Sun, Darawish, {De Luca}, Wang, Dumslaff, Narita, M{\"{u}}llen, Ruffieux,
  Meunier, Fasel, and Calame]{Overbeck2019UniversalLengthdependent}
Overbeck,~J.; Barin,~G.~B.; Daniels,~C.; Perrin,~M.~L.; Braun,~O.; Sun,~Q.;
  Darawish,~R.; {De Luca},~M.; Wang,~X.~Y.; Dumslaff,~T.; Narita,~A.;
  M{\"{u}}llen,~K.; Ruffieux,~P.; Meunier,~V.; Fasel,~R.; Calame,~M. A
  Universal Length-dependent Vibrational Mode in Graphene Nanoribbons.
  \emph{{ACS Nano}} \textbf{2019}, \emph{13}, 13083--13091\relax
\mciteBstWouldAddEndPuncttrue
\mciteSetBstMidEndSepPunct{\mcitedefaultmidpunct}
{\mcitedefaultendpunct}{\mcitedefaultseppunct}\relax
\EndOfBibitem
\bibitem[Koch \latin{et~al.}(2006)Koch, Von~Oppen, and Andreev]{koch2006theory}
Koch,~J.; Von~Oppen,~F.; Andreev,~A. Theory of the Franck-Condon blockade
  regime. \emph{Physical Review B} \textbf{2006}, \emph{74}, 205438\relax
\mciteBstWouldAddEndPuncttrue
\mciteSetBstMidEndSepPunct{\mcitedefaultmidpunct}
{\mcitedefaultendpunct}{\mcitedefaultseppunct}\relax
\EndOfBibitem
\bibitem[Soler \latin{et~al.}(2002)Soler, Artacho, Gale, Garc{\'{i}}a,
  Junquera, Ordej{\'{o}}n, and S{\'{a}}nchez-Portal]{Soler2002}
Soler,~J.~M.; Artacho,~E.; Gale,~J.~D.; Garc{\'{i}}a,~A.; Junquera,~J.;
  Ordej{\'{o}}n,~P.; S{\'{a}}nchez-Portal,~D. {The SIESTA method for ab initio
  order- N materials simulation}. \emph{Journal of Physics: Condensed Matter}
  \textbf{2002}, \emph{14}, 2745--2779\relax
\mciteBstWouldAddEndPuncttrue
\mciteSetBstMidEndSepPunct{\mcitedefaultmidpunct}
{\mcitedefaultendpunct}{\mcitedefaultseppunct}\relax
\EndOfBibitem
\bibitem[Sadeghi \latin{et~al.}(2015)Sadeghi, Sangtarash, and
  Lambert]{Sadeghi2015ElectronHeatTransport}
Sadeghi,~H.; Sangtarash,~S.; Lambert,~C.~J. Electron and Heat Transport in
  Porphyrin-based Single-molecule Transistors with Electro-burnt Graphene
  Electrodes. \emph{Beilstein Journal of Nanotechnology} \textbf{2015},
  \emph{6}, 1413--1420\relax
\mciteBstWouldAddEndPuncttrue
\mciteSetBstMidEndSepPunct{\mcitedefaultmidpunct}
{\mcitedefaultendpunct}{\mcitedefaultseppunct}\relax
\EndOfBibitem
\bibitem[Ferrer \latin{et~al.}(2014)Ferrer, Lambert, Garc{\'{i}}a-Su{\'{a}}rez,
  Manrique, Visontai, Oroszlany, Rodr{\'{i}}guez-Ferrad{\'{a}}s, Grace, Bailey,
  Gillemot, Sadeghi, and Algharagholy]{Ferrer2014}
Ferrer,~J.; Lambert,~C.~J.; Garc{\'{i}}a-Su{\'{a}}rez,~V.~M.; Manrique,~D.~Z.;
  Visontai,~D.; Oroszlany,~L.; Rodr{\'{i}}guez-Ferrad{\'{a}}s,~R.; Grace,~I.;
  Bailey,~S.~W.; Gillemot,~K.; Sadeghi,~H.; Algharagholy,~L.~A. {GOLLUM: A
  next-generation simulation tool for electron, thermal and spin transport}.
  \emph{New Journal of Physics} \textbf{2014}, \emph{16}, 1--37\relax
\mciteBstWouldAddEndPuncttrue
\mciteSetBstMidEndSepPunct{\mcitedefaultmidpunct}
{\mcitedefaultendpunct}{\mcitedefaultseppunct}\relax
\EndOfBibitem
\end{mcitethebibliography}

\end{document}